\documentclass[letterpaper,twocolumn,10pt]{article}
\usepackage{usenix-2020-09}
\usepackage{graphicx}
\usepackage{booktabs}
\usepackage{hyperref}
\usepackage{multirow}
\usepackage{makecell}
\usepackage{subcaption}
\usepackage{caption} 
\usepackage{algorithm}
\usepackage{algpseudocode}
\usepackage{amsmath}
\usepackage[shortlabels]{enumitem}
\usepackage[flushleft]{threeparttable}

\usepackage{tcolorbox}

\usepackage{listings}

\usepackage{tikz}
 
\usepackage{pifont}

\usepackage{tabularx}
\usepackage{xcolor}
\begin{document}

\title{The Cure is in the Cause: A Filesystem for Container Debloating}
\author{
{\rm Huaifeng Zhang}\\
Chalmers University of Technology
\and
{\rm Mohannad Alhanahnah}\\
University of Wisconsin-Madison
\and
{\rm Philipp Leitner}\\
Chalmers University of Technology
\and
{\rm Ahmed Ali-Eldin}\\
Chalmers University of Technology
} 

\maketitle

\begin{abstract}
    Containers have become a standard for deploying applications due to their convenience, but they often suffer from significant software bloat—unused files that inflate image sizes, increase provisioning times, and waste resources.
These inefficiencies are particularly problematic in serverless and edge computing scenarios, where resources are constrained, and performance is critical.
Existing debloating tools are limited in scope and effectiveness, failing to address the widespread issue of container bloat at scale.
In this paper, we conduct a large-scale evaluation of container bloat, analyzing the top 20 most downloaded containers on DockerHub.
We evaluate two state-of-the-art debloating tools, identify their limitations, and propose a novel solution, BAFFS, which addresses bloat at the filesystem level by introducing a flexible debloating layer that preserves the layered structure of container filesystems.
The debloating layer can be organized in different ways to meet diverse requirements.
Our evaluation demonstrates that over 50\% of the top-downloaded containers have more than 60\% bloat, and BAFFS reduces container sizes significantly while maintaining functionality.
For serverless functions, BAFFS reduces cold start latency by up to 68\%.
Additionally, when combined with lazy-loading snapshotters, BAFFS enhances provisioning efficiency, reducing conversion times by up to 93\% and provisioning times by up to 19\%.
\end{abstract}


\section{Introduction}

Container technology is one of the main deployment models for applications running machine learning~\cite{lee2018pretzel}, serverless computing~\cite{akkus2018sand}, or general computing~\cite{qiu2020firm} on cloud~\cite{anwar2018improving}, and edge systems~\cite{chen2022starlight}. Containers' popularity is driven by how the technology simplifies application deployments. To reduce storage requirements and increase network efficiency, containerized applications are distributed as container images, composed of compact, and shareable ``layers'' of files. These ``image'' and their ``layer'' are stored in centralized registries~\cite{anwar2018improving}, such as DockerHub~\cite{dockerhub}. A user ``pulls'' a container from a registry, and then can either update the container image, create a new container in the process, or use the pulled image as is.
Containers offer great convenience but at the expense of efficiency.
Containers usually package unnecessary files and their sizes have grown significantly over the years~\cite{wang2021faasnet, zhao2019large}.
In this paper, we argue that a large percentage of these containers is \emph{software bloat}, i.e., code and features that are never used during deployment.

Software bloat is a result of how software is developed today, with developers constantly adding features and including dependencies that are useful only to a subset of users while the rest is not (or seldom) used~\cite{zhang2022machine,agadakos2020large}. Bloat leads to reduced performance, wasted resources, and increased vulnerabilities~\cite{kuo2020set}.
Bloat exists in all software layers, from operating systems~\cite{kuo2020set}, to machine learning frameworks~\cite{zhang2022machine}, and web applications~\cite{webdebloat}.

When containers are used, the bloat problem is exacerbated due to their use of layered filesystems, such as
the union file-system~\cite{quigley2006unionfs},  the OverlayFS~\cite{overlayfs}, and  BTRFS~\cite{rodeh2013btrfs}, on which containers depend. In a layered filesystem, a container inherits all the filesystem layers of its \emph{base container image}. Any bloat in the base image is inherited in the new container, with new bloat possibly added with each layer added to the filesystem.
Software bloat poses significant challenges in modern computing technologies.
It wastes resources in cloud data centers and edge devices with limited resources~\cite{chen2022starlight,anwar2018improving}.
One of the most noticeable impacts of bloat is the increased cold start latency in serverless computing:
Bloat in containers leads to prolonged provisioning times for serverless functions, further exacerbating the cold start latency problem~\cite{anup2019agile,li2022help,vahidinia2020cold}

Container debloating aims to address the negative impacts of bloat by removing unnecessary components from container images.
Several container debloaters have been proposed to achieve this~\cite{Cimplifier,dockerslim,ghavamnia2020confine,lei2017speaker}.
A recent study ~\cite{hassan2023evaluating} evaluated three container debloaters on seven containers, demonstrating that the containers were significantly bloated.
However, the number of containers evaluated is too limited to draw general conclusions about the extent of bloat in containers.
Furthermore, the results show that these debloaters failed to debloat some containers.
Therefore, the applicability of these container debloaters to a broader set of containers remains unclear.
Additionally, the impact of debloating on the performance of real-world applications, such as serverless computing, has yet to be thoroughly investigated.

Related to the problem of bloat, the system community has focused on solving one of the key side effects of container bloat, reducing container provisioning times~\cite{chen2022starlight,fu2020fast,gu2023lopo}.
Techniques such as lazy-loading snapshotters~\cite{chen2022starlight, eStargz}, splitting containers into fat and slim images~\cite{thalheim2018cntr}, and developing new image storage drivers~\cite{harter2016slacker} have been proposed to improve provisioning efficiency.
However, these methods address the symptoms of the issue rather than the root cause, which is the underlying software bloat in containers.
Container debloating is orthogonal to these approaches and can be combined with them to achieve further performance enhancements.

To gain a broader understanding of (1) the general extent of bloat in containers and (2) the effectiveness of existing tools in debloating them, we propose a container debloating framework.
Using the framework, we analyze the top 20 most downloaded containers from DockerHub and test two state-of-the-art container debloaters.
The results reveal several intrinsic limitations in existing container debloating approaches.
Driven by these findings, we argue that bloat is best managed and removed at the filesystem level.
We introduce BAFFS, a Bloat-Aware Flexible Filesystem, a novel filesystem-level container debloating approach that overcomes the limitations of existing container debloaters.
Our contributions are as follows:
\begin{itemize}
    \item We propose a framework to debloat containers and evaluate the effectiveness of container debloaters. We analyze the top 20 most downloaded containers from DockerHub, evaluate two state-of-the-art debloaters, and identify their limitations. The framework is open-sourced at To Be Added.
    \item We introduce BAFFS, a novel filesystem-level debloating approach that overcomes the limitations of existing container debloaters. We apply BAFFS to successfully debloat all 20 containers and show that more than 50\% of the containers have more than 60\% bloat. BAFFS is open-sourced at \url{https://github.com/jzh18/BAFFS}.
    \item  We evaluate BAFFS using real-world serverless computing benchmarks and demonstrate that debloating can reduce the cold start latency of serverless functions by up to 68\%.
    \item We show that container debloating can be effectively combined with other container optimization techniques, such as lazy-loading snapshotters to enhance their performance, significantly reducing conversion time by up to 93\% and provisioning time by up to 19\%.
\end{itemize}

\section{Background}\label{sec:background}
In this section, we first describe serverless computing, which leverages containers to deploy serverless functions.
Then we introduce container filesystems, container debloating and lazy-loading snapshotters.

\subsection{Serverless Computing}

Serverless computing is an event-driven execution model in which the cloud provider dynamically manages the allocation and scaling of resources.
Developers focus solely on writing and deploying discrete units of functionality, commonly referred to as serverless functions, without needing to manage the underlying infrastructure.
This design allows developers to break down applications into small, manageable units of code that are easy to scale, maintain, and reuse across various workflows.
Serverless functions are stateless, event-triggered, and designed to execute well-defined computation tasks, making them highly modular~\cite{li2022serverless,bilal2023great}.
Containers are widely used for deploying serverless functions, providing an isolated and consistent environment for function execution~\cite{djemame2020open,githubGitHubOpenfaasfaas,amazonServerlessFunction}.

Despite their benefits, serverless computing faces significant performance challenges, particularly during the cold start phase.
A cold start occurs when a serverless function is invoked for the first time, or after a period of inactivity, requiring the serverless platform to provision the execution environment.
This process can involve downloading the container image from the registry, initializing the runtime environment, and executing the function, all of which contribute to latency~\cite{anup2019agile,li2022help,vahidinia2020cold}.
Serverless function containers are bloated, which further exacerbates the cold start latency problem.
For instance, on AWS Lambda, a widely used serverless computing platform~\cite{amazonServerlessFunction}, a simple serverless function that prints "Hello World" can require a container image as large as 530 MB~\cite{amazonDeployPython}.

\subsection{Container File Systems}\label{sec:container_fs}

LXC~\cite{lxc}, and Docker~\cite{docker} are the two most widely used technologies for containerization.
In essence, a container is a process running on the host operating system isolated via operating system primitives such as cgroups, namespaces, and typically using a union filesystem such as OverlayFS~\cite{overlayfs} and BTRFS~\cite{rodeh2013btrfs} or any similar approaches~\cite{sun2020baoverlay,zheng2018wharf}.

\begin{figure}[htbp]
  \centering
  \includegraphics[scale=0.36]{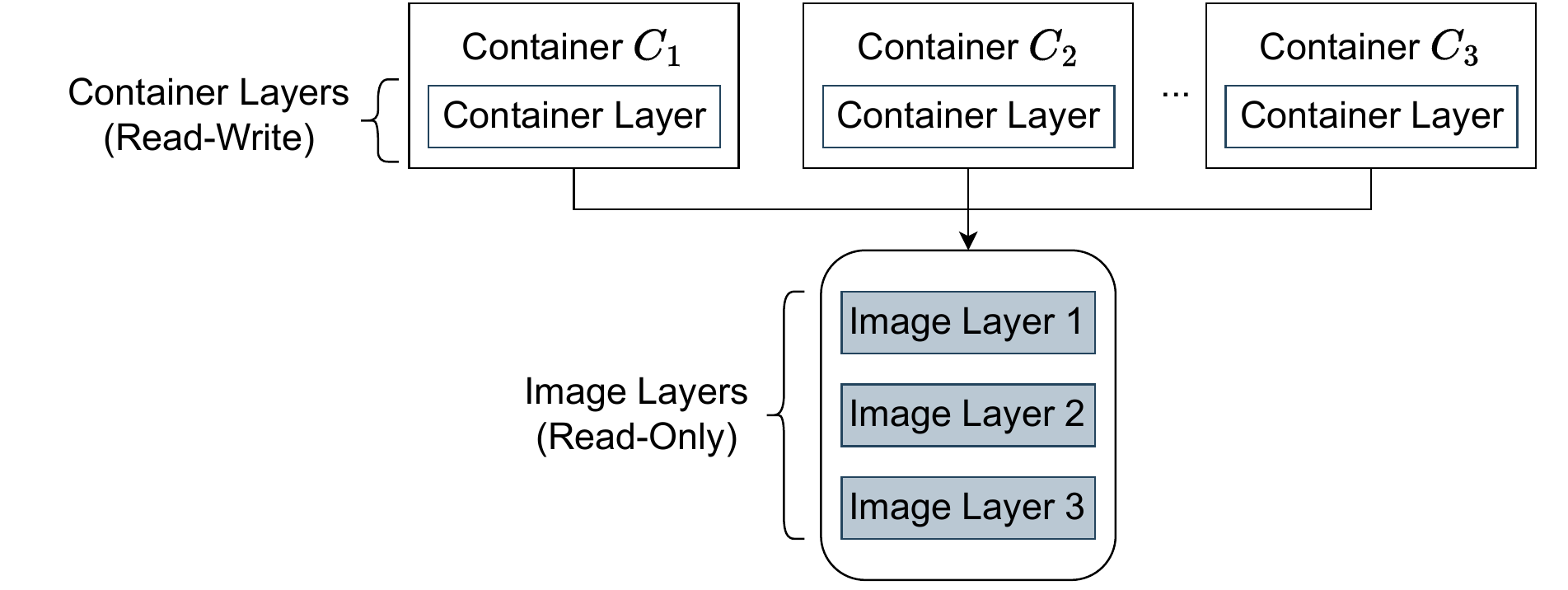}
  \caption{An example of a container filesystem.}
  \label{fig:container_fs}
  \vspace{-1.36em}
\end{figure}

A container filesystem can be thought of as a \emph{container layer} and a list of \emph{image layer}s, as illustrated in Figure~\ref{fig:container_fs}.
The container layer handles all write requests from the container in a copy-on-write manner. All files created or modified by the container are stored in the container layer.
The image layers are read-only.
When a container accesses a file, the filesystem first looks up the file in the container layer, then in the image layers one by one, from the top to the bottom, until the file is found, or an error is raised.

The image layers can be shared across many containers, with the container layer allowing each container to have its isolated filesystem.
This approach greatly reduces the disk space required for each container. However, this layered filesystem architecture is also a major source of bloat\cite{Cimplifier,thalheim2018cntr,dockerslim}. Most containers are built on top of a base container image, where a container inherits all the filesystem layers of the base container image. Any bloat in the base image is inherited in the new container, with new bloat possibly added with each layer added to the filesystem.

\subsection{Container Debloating}

Bloated containers can consume significant disk space and network bandwidth, which can slow down the provisioning performance of containerized applications~\cite{zhang2022machine,hu2023cec,sangolli2019enabling}, e.g., long cold start times for serverless functions~\cite{anup2019agile,li2022help,vahidinia2020cold}.
Moreover, large containers often include unnecessary dependencies, libraries, and applications that can introduce security vulnerabilities.
Container debloating removes bloat from container image layers to reduce its size and improve its security and performance.
Cimplifier is an example of a container debloater~\cite{Cimplifier}. Cimplifier uses dynamic analysis to slim and partition containers to provide better privilege separation across applications. SlimToolKit~\cite{dockerslim} is another open-source debloater that also relies on dynamic analysis to generate \emph{Seccomp} profiles and remove unnecessary files from containers.
Another category of container debloating does not aim to reduce the size of containers but instead focuses on restricting the system calls available to containers~\cite{ghavamnia2020confine,lei2017speaker}, thus reducing the attack surface of the container.

Size-oriented container debloating, such as Cimplifier and SlimToolKit,  has the advantage of reducing the provisioning time of containers, while reducing the bandwidth requirements of the deployment, making deployments faster and more efficient, while saving space at the host.
However, these debloating tools are only tested using a limited number of containers and fail to debloat some containers~\cite{hassan2023evaluating,zhang2022machine}.
Their applicability to a broader range of containers is not well understood.

\subsection{Lazy Loading Snapshotters}
Another major direction to improve container provisioning time is the use of the snapshotter plugin~\cite{containerd} available with \emph{containerd} plugins to enable lazy-loading of containers. 
Snapshotter techniques need to convert the original container image to the snapshotter format.
Starlight~\cite{chen2022starlight} is an accelerator for container provisioning that redesigns the container provisioning protocol, filesystem, and storage format.
Starlight initially identifies the essential files needed for container start-up. When a container is pulled, these essential files are the first to be retrieved, allowing for a quick start-up.
The remaining files are then pulled in the background, even if they are not required for the container at runtime. As a result, Starlight finally pulls a fully bloated container on the device.
Consequently, Starlight does not remove bloat from the container.

Another snapshotter-based technique is eStargz~\cite{eStargz}. eStargz estimates the order of use of files in the image layers of a container and presorts the files in
each compressed layer according to the estimation. When a container starts and
opens a file that has not been transferred yet, eStargz pauses the container start-up and requests the file from the registry.
Similar to Starlight, the remaining files of the container are pulled in the background, resulting in a fully bloated container.

Container debloating and lazy-loading snapshotters are orthogonal techniques that can be effectively combined to further enhance the performance of containerized applications.

\section{Motivation}\label{sec:motivation}
In this section, we conduct a large-scale evaluation of bloat in the most popular containers on DockerHub and evaluate the effectiveness of two state-of-the-art container debloaters.
We analyze the debloating results and identify key limitations in the debloaters.
These findings motivate us to propose a new, more effective container debloating method.

\subsection{Container Debloating Evaluation Framework}
\begin{figure}[htbp]
  \centering
  \includegraphics[scale=0.6]{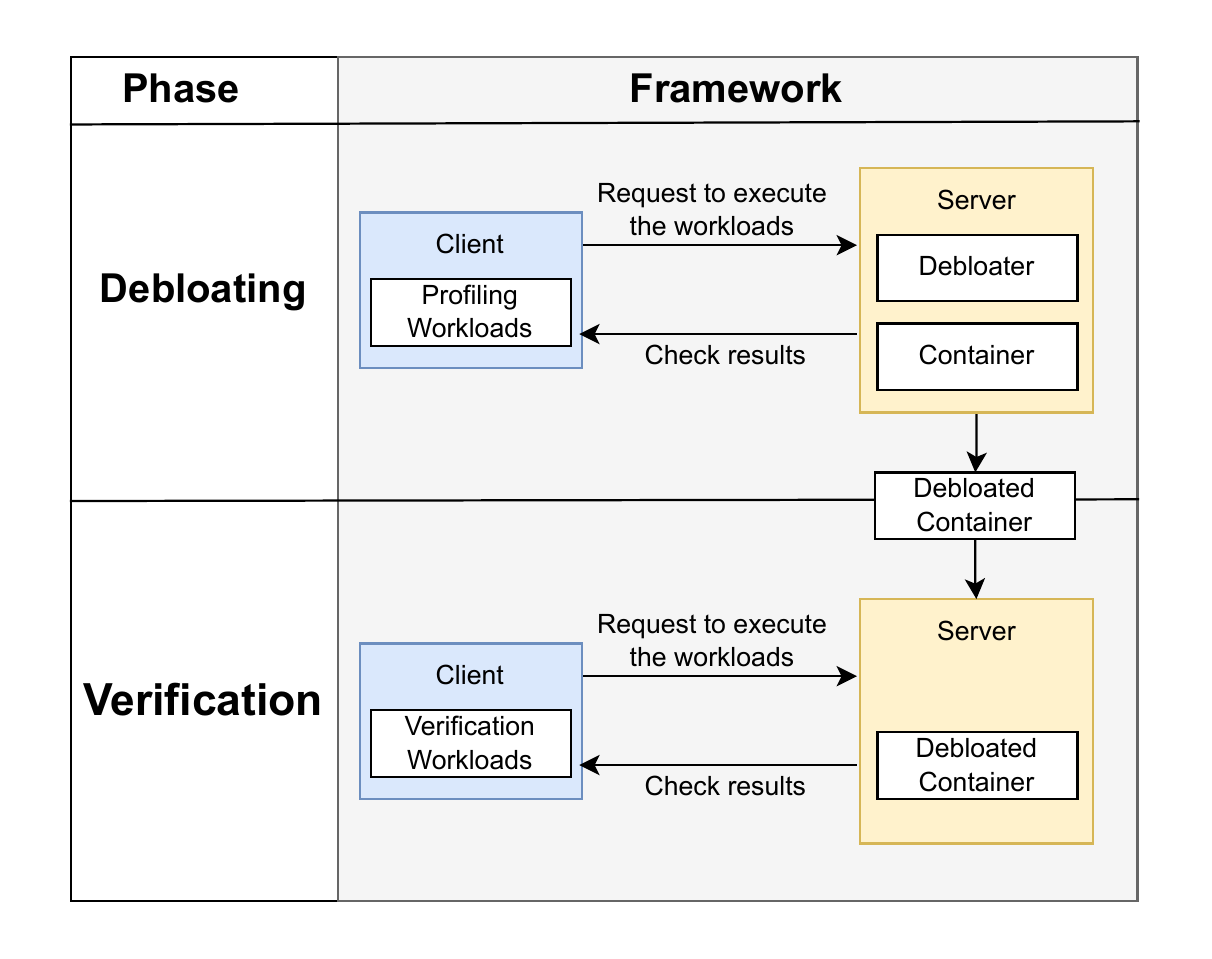}
  \caption{Container debloating evaluation framework.}
  \label{fig:debloater_eval_framework}
  \vspace{-1.36em}
\end{figure}

To facilitate a large-scale evaluation of container bloat, we developed a container debloating evaluation framework, as illustrated in Figure~\ref{fig:debloater_eval_framework}.
The framework has a client-server architecture.
The client is responsible for requesting the server to execute workloads and verify the execution results, while the server sets up the container and debloater environment, executes workloads within the container, and generates the debloated container.

Containers can be broadly categorized into service containers and command-line tool containers.
Service containers, which are long-running, designed to start a service and handle requests, fit the client-server architecture naturally,
For command-line tool containers, which are typically short-lived, executing a single task and exiting, the server of the framework starts them as long-running containers, and the client of the framework utilizes the containers' \texttt{exec} feature to run workloads within the containers~\cite{dockerDockerExec}.
In doing so, the client-server architecture enables the framework to provide a unified interface for handling both types of containers effectively.
We develop a framework similar to the binary debloating evaluation framework developed by Brown et al.~\cite{brown2024broad}.

The framework operates in two phases: debloating and verification.
During the debloating phase, the server executes profiling workloads and generates a debloated container based on those workloads.
The verification phase verifies whether the debloated containers can execute the verification workloads successfully.
Note that the verification workloads can be different or the same as the profiling workloads.
The profiling and verification workloads are structured as a list of test cases in Python.
Listing~\ref{lst:feature_code} presents a snippet of the profiling workloads for a \texttt{nginx} container, including workloads such as serving static content and proxy pass, with each workload implemented as an individual test case.

\begin{lstlisting}[language=Python,label={lst:feature_code},caption={Profiling workloads for a \texttt{nginx} container.},numbers=none]
class NginxProfilingWorkloads(unittest.TestCase):
  def test_static_content(self):
    response = requests.get(server_url)
    self.assertEqual(response.status_code, 200)
    self.assertIn("Hello, NGINX!", response.text)
  ...#other test cases are omitted for brevity
  def test_proxy_pass(self):
    response = requests.get(proxy_url)
    self.assertEqual(response.status_code, 200)
    self.assertIn("Hello, Proxy!", response.text)
\end{lstlisting}

\subsection{Top 20 Downloaded Containers Bloat}
We select the top 20 downloaded containers from DockerHub, focusing exclusively on application containers rather than general-purpose operating system containers.
The latter are not designed for specific applications and are hence unsuitable for debloating.
For each selected container, we manually identify its representative workloads to cover as many relevant use-cases as possible.
The number of identified workloads for each container ranges from 2 to 7.
Each workload may include multiple functionalities of the container.
The details of the selected containers are shown in Table~\ref{tab:containers} in the appendix.
We select Cimplifier and SlimToolKit, to debloat these containers.
While tools such as Confine~\cite{ghavamnia2020confine} and Speaker~\cite{lei2017speaker} exist, they do not remove bloat from containers and are thus excluded from our evaluation.

\textbf{Size reduction}. Table~\ref{tab:cimplifier-dockerslim-results} summarizes the debloating results for successfully debloated containers by Cimplifier and SlimToolKit.
A container is considered successfully debloated if it can execute all the profiling workloads without errors.
Cimplifier and SlimToolKit successfully debloated 7 and 8 containers, respectively.
The results also reveal significant bloat across these containers, with average size reductions of 51\% for Cimplifier and 55\% for SlimToolKit.
These highly bloated containers further underscore the need to analyze and address bloat in these widely used containers.
The details of the debloating results for all containers are provided in Tables~\ref{tab:cimplifier_all_size_reduction} and \ref{tab:slimtool_all_size_reduction} in the appendix.

\begin{table}[h]
  \centering
  \caption{Number of successfully debloated containers and the average, minimum, and maximum size reduction.}
  \label{tab:cimplifier-dockerslim-results}
  \begin{tabular}{lrrrr}
    \toprule
    Debloater   & \#Containers & Avg. & Min.  & Max. \\
    \midrule
    Cimplifier  & 7            & 51\% & 4\%   & 95\% \\
    SlimToolKit & 8            & 55\% & 4\% & 93\% \\
    \bottomrule
  \end{tabular}
\end{table}

\textbf{Effectiveness}. We also analyze the limitations of the two debloaters, both of which failed to successfully debloat more than 50\% of the containers.
By examining their underlying mechanisms, we identified several reasons for these failures.
Using tracing techniques such as \texttt{ptrace} or \texttt{strace}~\cite{man7Ptrace2Linux,man7Strace1Linux}, Cimplifier and SlimToolKit monitor files accessed by the profiling workloads and remove files not accessed.
However, this approach has the following key limitations:
\begin{itemize}
  \item The tracing techniques may fail to capture certain file accesses. For instance, a file accessed through commands like \texttt{chown user filename} is not detected by these techniques, resulting in the removal of used files.
  \item Both debloaters only begin tracing after the container has started running. Hence, they cannot capture files accessed during the startup phase. This is especially problematic for containers that utilize resources at startup, such as CUDA libraries in machine learning containers. Removal of these files can lead to the container failing to start.
  \item Both debloaters break the layer structure of the container filesystem, thereby negating the advantages of layer sharing.
        For example, the total size of debloated containers can be larger than the original containers due to the broken layer structure.
\end{itemize}
While the first limitation could be addressed with further development, the last two limitations are inherent.
These fundamental issues limit the effectiveness of current container debloating methods.

\textbf{Privacy}. An additional issue observed with SlimToolKit is that the debloated container can, in some cases, become larger than the original container.
This occurs because SlimToolKit commits changes made during profiling to the container, altering its state.
For example, in the case of the \texttt{mongodb} container, the profiling workloads insert data into the database, and the resulting debloated container retains that data.
This could lead to potential privacy leaks if the debloated container is published to a public registry.
Similarly, in other containers, profiling workloads may write files to the container.
These files are then captured by the SlimToolKit and retained in the debloated container, resulting in a debloated container that is larger than its original version.

\subsection{Summary}

Our findings reveal that existing container debloaters are not effective at debloating containers and face inherent limitations.
Additionally, the significant size reductions observed in several successfully debloated containers indicate that popular containers are highly bloated.
Currently, no viable solution exists for container debloating that is suitable for production systems.
Meanwhile, there is a growing trend of building extremely large container images, such as those used in machine learning, which can exceed tens of gigabytes~\cite{zhang2022machine}.

To address the challenges of container debloating, we believe that the problem is best tackled at the filesystem level.
We propose a novel container debloater that operates at the filesystem level by leveraging the layered structure of container filesystems.
The details of our approach are presented in the next section.

\section{BAFFS: A Bloat Aware File-System}

This section introduces BAFFS, a bloat-aware flexible filesystem that debloats containers.
BAFFS has a basic building block called \textit{debloating layer}, which is used to detect used files.
BAFFS offers greater flexibility compared to existing container debloaters.
By combining debloating layers with original image layers in different ways, BAFFS can operate in different modes.
We present three modes of BAFFS: no-sharing, fully-sharing, and semi-sharing, which are suitable for different scenarios.

\subsection{Debloating Layer}\label{sec:debloating_layer_reloading_layer}
A container filesystem includes a list of image layers as described in \S~\ref{sec:container_fs}.
BAFFS introduces a novel layer, the debloating layer, to the original container filesystem.
Algorithm~\ref{alg:blafs_read} shows the pseudocode of reading a file from a BAFFS filesystem.
A layer $L$ can be considered a directory.
BAFFS tries to open the file layer by layer as the original filesystem.
The difference is that BAFFS calls DOPEN to open the file from a debloating layer.
We next introduce the debloating layer in detail.

\begin{algorithm}[htbp]
    \caption{Read up to $c$ bytes from path $p$ into buffer $buf$.}\label{alg:blafs_read}
    \begin{algorithmic}[1]
        \Require{File path $p$; buffer $buf$; count $c$}
        \Ensure{Error code}
        \For{$L$ in $filesystem$} \Comment{from top to bottom}
        \If{$L$ is a debloating layer}
        \State $fd \gets$ DOPEN($L$,$p$)
        \Else
        \State $fd \gets$\texttt{open}($p$)
        \EndIf
        \If{$fd\not\equiv{-1}$}
        \State \Return \texttt{read}($fd$,$buf$,$c$)
        \EndIf
        \EndFor
        \State \Return $-1$
    \end{algorithmic}
\end{algorithm}

Debloating layers are used to detect used files.
A debloating layer $D$ has a list of image layers as its child layers ($D.children$).
When reading a file from a debloating layer, the file is moved from the child layers to the debloating layer.
In doing so, the debloating layer only includes used files.
Algorithm \ref{alg:debloat_file_req} shows details of this process.
The file to be read is $p$.
We use the notation $D$/$p$ to denote concatenating the path of layer $D$ with path $p$.
The algorithm first opens $p$ in $D$. If $p$ exists in $D$ ($fd\not\equiv{-1}$),  it returns its file descriptor.
Otherwise, BAFFS tries to open $p$ from the child layers.
If $p$ exists in layer $l$, the file is moved from $l$ to $D$.
If $p$ does not exist in the child layers, -1 is returned to indicate that the file does not exist in $D$.

\begin{algorithm}[htbp]
    \caption{Pseudo code of function DOPEN}\label{alg:debloat_file_req}
    \begin{algorithmic}[1]
        \Require{Debloating layer $D$; file path $p$}
        \Ensure{Error code}
        \Function{dopen}{$D$,$p$}
        \State $fd \gets$ \texttt{open}($D$/$p$)
        \If{$fd\not\equiv{-1}$}
        \State \Return $fd$
        \EndIf
        \For{$l$ in $D.children$} \Comment{from top to bottom}
        \State $fd \gets$ \texttt{open}($l$/$p$)
        \If{$fd\not\equiv{-1}$}
        \State \texttt{move($l$/$p$,$D$/$p$)}
        \State \Return \texttt{open}($D$/$p$)
        \EndIf
        \EndFor
        \State \Return $-1$
        \EndFunction
    \end{algorithmic}
\end{algorithm}

\begin{figure}[htbp]
    \centering
    \includegraphics[scale=0.5]{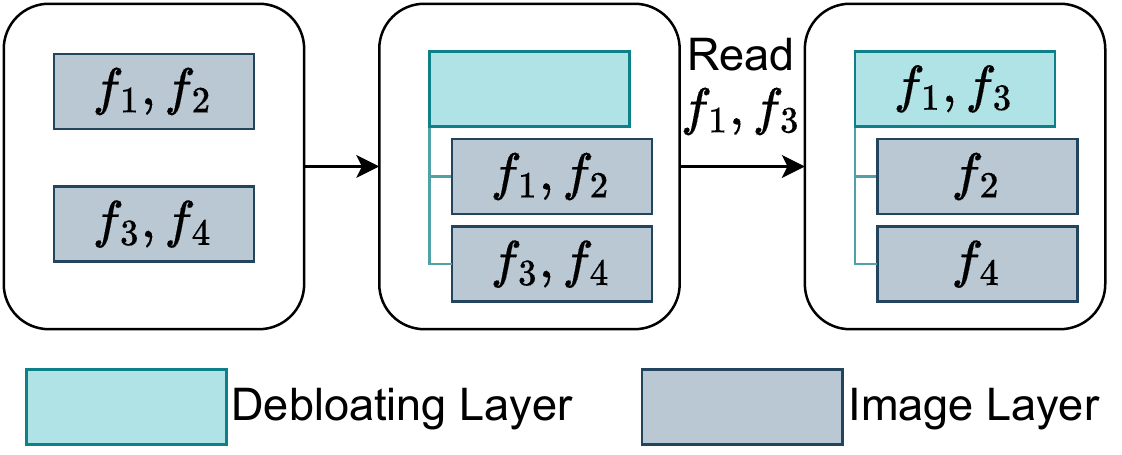}
    \caption{An example of debloating layers.}
    \label{fig:debloating_layer_example}
\end{figure}

While the original container filesystem is linear-layered where the layers are stacked sequentially, BAFFS is a tree-layered filesystem due to the debloating layers' child layers.
It is noteworthy that line 1 in Algorithm~\ref{alg:blafs_read} only iterates the root layers (layers that do not have parent layers), the child layers are not iterated.
Figure~\ref{fig:debloating_layer_example} shows an example of the mechanism of debloating layers.
The original filesystem of the container has two image layers and each layer has two files.
Then a debloating layer is added to the filesystem.
The debloating layer has two child layers, which are the two image layers.
The debloating layer is the only root layer in the filesystem.
When the container reads files $f_1$ and $f_3$, the two files are moved to the debloating layer from the image layers.
Finally, the two used files are in the debloating layer, while the other two unused files are in the image layers.
If we remove the image layers, the container will be debloated and only contains the two used files.

\subsection{Three Modes of BAFFS}\label{sec:debloating_phase}
Debloating a container using BAFFS involves three steps: \emph{Convert}, \emph{Profile} and \emph{Export}.
The convert step converts the original container filesystem to a BAFFS filesystem by adding debloating layers.
The profile step has the converted container performing some profiling workloads.
The files used by the profiling workloads are moved from image layers to debloating layers.
In doing so, used files and unused files are identified and separated.
The export step removes all the image layers from the container and generates a debloated container.

At the convert step, BAFFS can be converted to different modes by adding debloating layers in different ways.
Here we present three modes of BAFFS: \textit{no-sharing}, \textit{fully-sharing}, and \textit{semi-sharing}.
Operating in the no-sharing mode results in a container with maximum per-container size reduction;
while operating in the fully-sharing mode results in multiple containers with maximum total size reduction;
the semi-sharing mode is designed for serverless computing.
We now detail the three modes.

\subsubsection{No-Sharing and Fully-Sharing Modes}
\textbf{No-sharing Mode}. In this mode, BAFFS guarantees the minimum possible size of a container for the profiling workloads by removing files that are not used by the profiling workloads.
Given a container with $n$ image layers, the no-sharing BAFFS inserts one debloating layer, which has all the $n$ image layers as its child layers.
The debloated container contains only files needed for profiling workloads.
This mode is equivalent to existing container debloaters, such as Cimplifier and SlimToolKit.

While the no-sharing BAFFS is effective in producing debloated containers with a minimum size per container, duplicate files among multiple debloated containers that share the same host can result in the total size of the debloated containers being larger than the total size of the original ones which share filesystem layers.
We note that all existing container debloaters suffer from this flaw.

\begin{figure}
    \centering
    \includegraphics[scale=0.44]{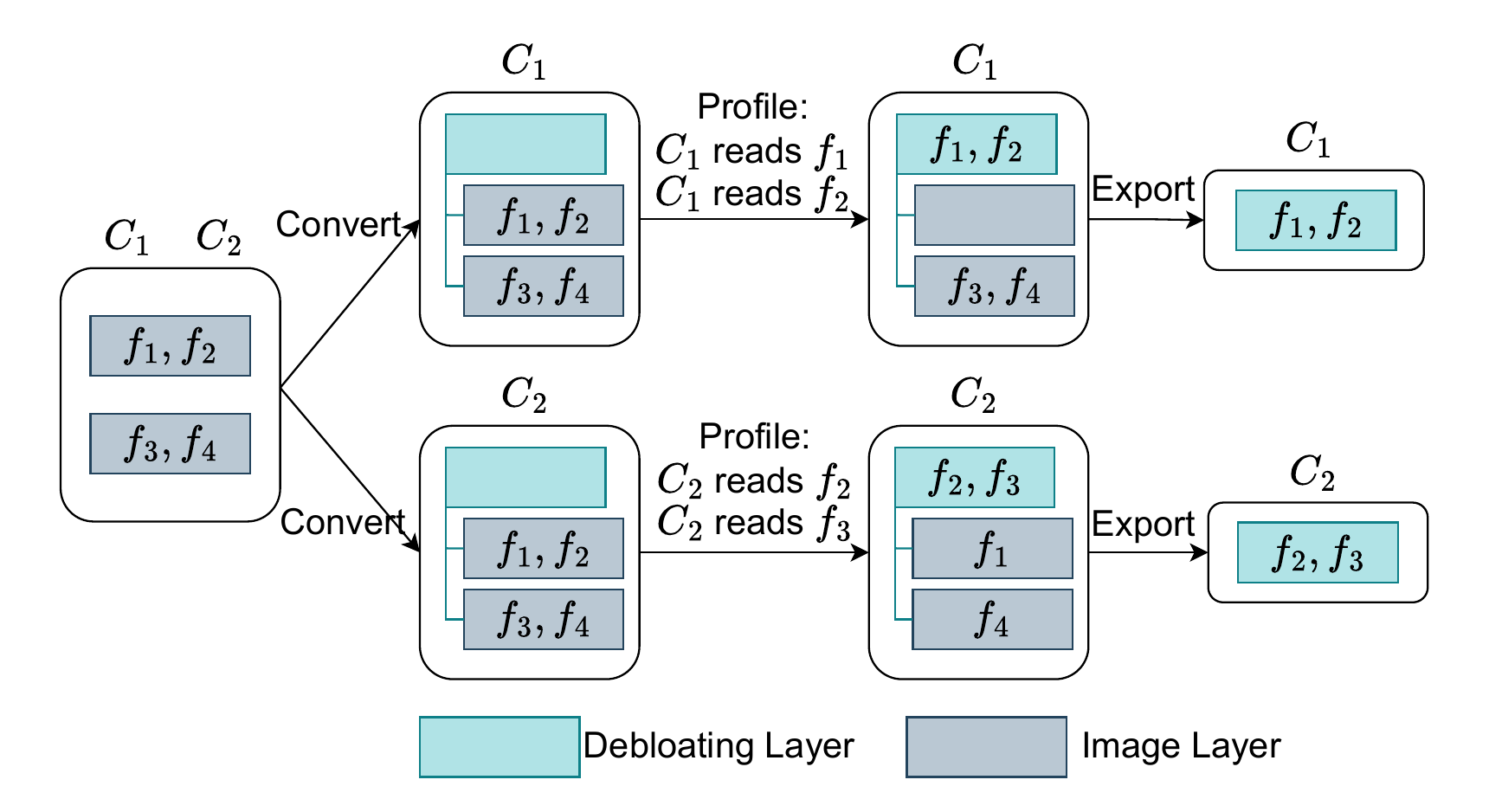}
    \caption{Example for no-sharing BAFFS}
    \label{fig:no_share_example}
    \vspace{-1.26em}
\end{figure}

\textbf{Example for the no-sharing mode}. Figure \ref{fig:no_share_example} displays an example of no-sharing BAFFS.
The two containers, $C_1$ and $C_2$, share the same image layers, which include two layers and four files $f_1$, $f_2$, $f_3$ and $f_4$.
The sizes of $f_1$, $f_2$, $f_3$, and $f_4$ are 1MB, 2MB, 3MB, and 4MB respectively.
So, the sizes of $C_1$ and $C_2$ are both 10MB.
First, the image layers are converted to BAFFS by prepending a debloating layer for $C_1$ and $C_2$ respectively.
The debloating layers have the two image layers as their child layers.
When profiling, $C_1$ reads $f_1$ and $f_2$, so $f_1$ and $f_2$ are moved to the debloating layer from the image layer.
Similarly, $f_2$ and $f_3$ are moved to the debloating layer of $C_2$ from the image layers.
After exporting, the debloated version of $C_1$ has one debloating layer, which contains $f_1$ and $f_2$.
The debloated version of $C_2$ has one debloating layer, containing $f_2$ and $f_3$.
As a result, the sizes of $C_1$ and $C_2$ are reduced to 3MB and 5MB, which are 70\% reduction and 50\% reduction respectively.
However, the total size of debloated $C_1$ and $C_2$ is 8MB, which is only a 20\% reduction.
$f_2$ is duplicated in both debloated containers.

\textbf{Fully-sharing mode:} This mode allows for the sharing of common files across multiple debloated containers located on the same host.
Given a container with $n$ image layers, the fully-sharing BAFFS inserts $n$ debloating layers, each of which has one image layer as its child layer.
Compared with the no-sharing mode, this can lead to a larger size for each container, but the total storage used is reduced since the layers are shared.

\begin{figure}
    \centering
    \includegraphics[scale=0.42]{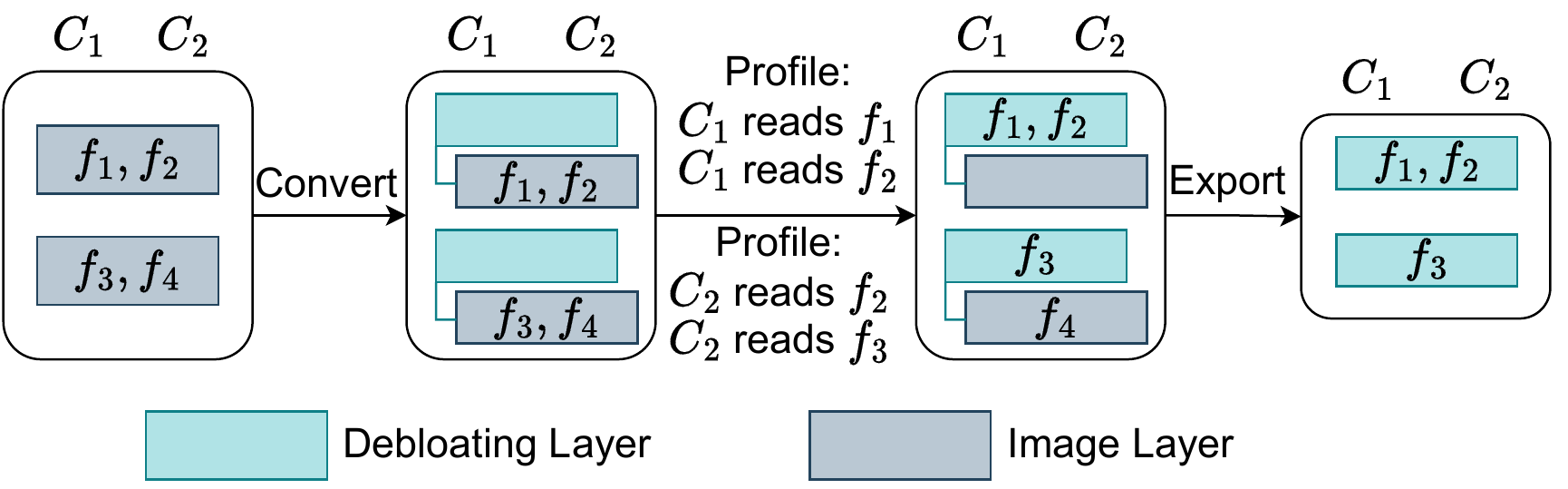}
    \caption{Example for fully-sharing BAFFS}
    \label{fig:share_example}
    \vspace{-1.36em}
\end{figure}

\textbf{Example for the fully-sharing mode}. An example of BAFFS with fully-sharing is illustrated in Figure \ref{fig:share_example}.
The input container $C_1$ and $C_2$ are the same as in Figure \ref{fig:no_share_example}.
First, two debloating layers are inserted having each image layer as their child layers.
The inserted debloating layers are shared by $C_1$ and $C_2$.
The same profiling workloads as in Figure~\ref{fig:no_share_example} are performed for $C_1$ and $C_2$ respectively.
After exporting, the debloated versions of $C_1$ and $C_2$ share the two debloating layers.
Therefore, the sizes of $C_1$ and $C_2$ are the same, which is 6MB.
The size is larger than the sizes of 3MB and 5MB generated by the no-sharing BAFFS, because each container consists of an unnecessary file ($f_3$ for $C_1$,$f_1$ for $C_2$).
However, unlike no-sharing BAFFS, the debloating layers are shared by the two debloated containers.
So the total size of $C_1$ and $C_2$ is 6MB, which is smaller than the size of 8MB generated by no-sharing BAFFS.

\textbf{Mode selection strategy}. To choose between no-sharing or fully-sharing, we implement a heuristic-based strategy that determines which mode to use.
Given $n$ containers $C_1$, $C_2$, $\cdots$, $C_n$.
Assume that these containers are debloated using the no-sharing BAFFS.
The resulting sizes of the debloated containers are $s_1$, $s_2$, $\cdots$, $s_n$ respectively.
The total disk size occupied, namely the \textbf{total size}, of the debloated containers is $t$.
We then compute the sizes of these containers if debloated with the fully-sharing BAFFS, $s'_1$, $s'_2$, $\cdots$, $s'_n$, respectively.
The total size of the debloated containers in this mode is $t'$.
Let $\alpha$ denote the size of duplicated files in $s_i$  because of no-sharing;
$\beta$ denotes the size of unnecessary files in $s'_i$ because of fully-sharing.
$\alpha$ is calculated by Equation \ref{eq:alpha},
\begin{equation}
    \label{eq:alpha}
    \alpha= \sum\nolimits_{1}^{n} s_i-t'
\end{equation}
$\beta$ can be obtained easily by Equation \ref{eq:beta},
\begin{equation}
    \label{eq:beta}
    \beta=\sum\nolimits_{1}^{n}(s'_i-s_i)
\end{equation}
Here we define $\theta$ as shown in Equation \ref{eq:theta},
\begin{equation}
    \label{eq:theta}
    \theta=\frac{\alpha}{\beta}=\frac{\sum\nolimits_{1}^{n} s_i-t'}{\sum\nolimits_{1}^{n}(s'_i-s_i)}
\end{equation}
A small number, e.g. $0.001$, can be added to the denominator to avoid a zero-by-zero divide error.
If $\theta$ is small, it means there are very few files that can be shared between the debloated containers ($\alpha$ is small), and many unnecessary files will be packed with each container if fully-sharing($\beta$ is large).
If $\theta$ is large, then many files can be shared between the debloated containers ($\alpha$ is large);
and fewer unnecessary files will be incurred if using the fully-sharing BAFFS ($\beta$ is small).
In our deployments, we set $1$ as a threshold for $\theta$.
No-sharing BAFFS is more appropriate for debloating if $\theta<1$;
otherwise, fully-sharing BAFFS is recommended.

\subsubsection{Semi-sharing: A Mode for Serverless Computing}
The semi-sharing mode is specifically tailored for serverless computing.
In serverless computing, a common approach to creating functions involves building a container on top of a base image, which is provided by the serverless platform~\cite{amazonServerlessFunction,djemame2020open}.
These base images are pre-existing in the serverless platform and are reused across multiple function containers.
The serverless platform only needs to pull the unique layers of the container, while reusing the layers derived from the base image.
This mechanism significantly reduces the time required for the cold start of a function.
The semi-sharing mode is explicitly designed for this scenario.

\begin{figure}[htbp]
    \centering
    \includegraphics[scale=0.4]{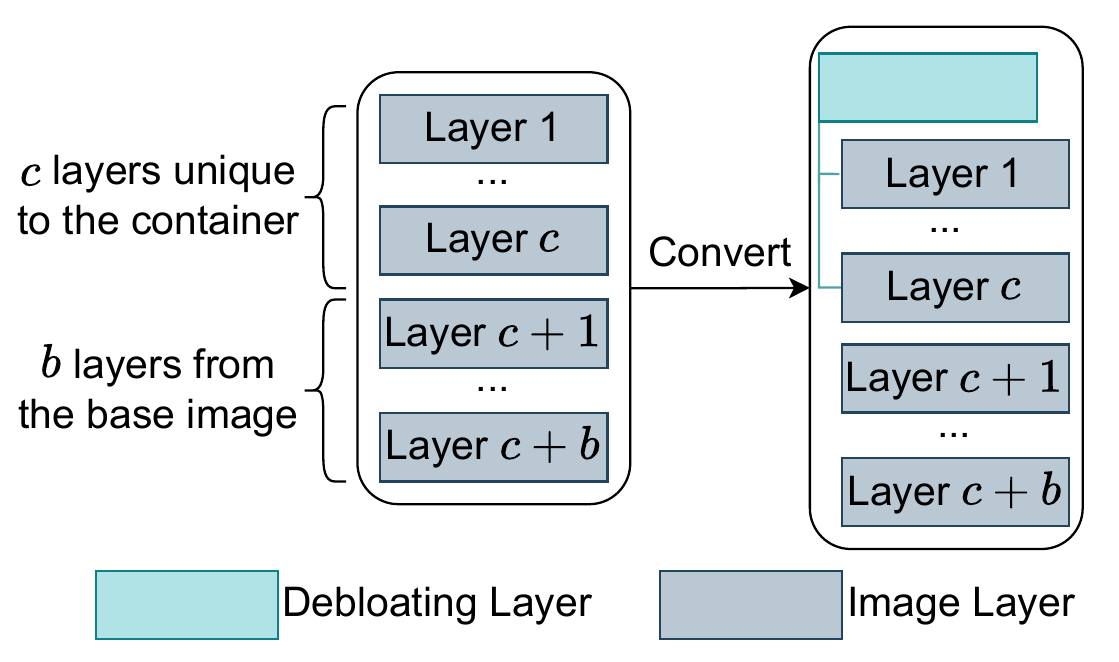}
    \caption{Convert the original container filesystem to the semi-sharing BAFFS filesystem.}
    \label{fig:semi-sharing}
    \vspace{-2em}
\end{figure}

The layers of a serverless function container comprise two segments: the top $c$ layers, which are unique to the container, and the bottom $b$ layers, which are shared with the base image.
The semi-sharing BAFFS inserts a single debloating layer, which has the $c$ unique layers as its child layers, and the layers from the base image remain unchanged, as shown in Figure \ref{fig:semi-sharing}.
Under this mode, only the unique layers of the container will be debloated, while the layers shared with the base image remain untouched.
This mode ensures that the debloated container retains the same base image as the original container, thereby can be reused by the serverless platform.
Consequently, the debloated container benefits from both a reduced size and the reuse of the base image.

\section{Evaluation}
We first evaluate BAFFS on the top 20 most downloaded containers from DockerHub.
Then we evaluate BAFFS on a serverless function benchmark to show the impacts of debloating on serverless functions.
Subsequently, we show how BAFFS improves the performance of lazy-loading snapshotters.
Finally, we evaluate the mode selection strategy of BAFFS.

\subsection{Evaluation on the Top 20 Containers}
\textbf{Effectiveness}.
Similar to the experiments presented in \S\ref{sec:motivation}, we debloated and verified the containers using the same workloads.
BAFFS successfully debloated all the 20 containers.
Table~\ref{tab:top_size_reduction} summarizes the debloating results for the top 10 containers, sorted by the reduction percentage.
The full results for all 20 containers are shown in Table~\ref{tab:all_size_reduction} in the appendix.
The original container sizes ranged from 14 MB to 1,241 MB, with size reductions varying between 4\% and 95\%.
Notably, more than half of the containers achieved a size reduction of over 60\%.
We also analyzed the correlation between the original container size and the size reduction, which yielded a value of -0.03, indicating that larger container sizes do not necessarily correspond to greater size reductions.

These findings show the effectiveness of BAFFS in debloating containers.
It also shows the significant bloat present in these widely-used containers.
Such bloat not only increases storage requirements but also amplifies network bandwidth usage and slows the time needed to pull or push containers to and from registries.

\begin{table}[h]
    \centering
    \begin{threeparttable}
        \caption{Debloating results for the top 10 containers sorted by the reduction percentage.}
        \label{tab:top_size_reduction}
        \begin{small}
            \begin{tabular}{lrrr}
                \toprule
                Container        & \makecell{Original              \\(MB)} & \makecell{Debloated\\(MB)} & Reduction \\
                \midrule
                httpd:2.4        & 141                & 7   & 95\% \\
                nginx:1.27.2     & 183                & 12  & 93\% \\
                memcached:1.6.32 & 81                 & 9   & 89\% \\
                mysql:9.1        & 574                & 99  & 83\% \\
                postgres:17      & 415                & 85  & 79\% \\
                ghost:5.101.3    & 547                & 121 & 78\% \\
                redis:7.4.1      & 112                & 27  & 75\% \\
                haproxy:3.0.6    & 98                 & 27  & 72\% \\
                mongo:8.0        & 815                & 233 & 71\% \\
                solr:9.7.0       & 561                & 195 & 65\% \\
                \bottomrule
            \end{tabular}
        \end{small}
    \end{threeparttable}
    \vspace{-1em}
\end{table}

\textbf{Generality}.
A common question in container debloating pertains to the generality of debloated containers: if a debloated container is used for a different task, will it still function correctly?
To answer this question, we debloated each container using only a single workload, deliberately selecting the most basic workload for each container to debloat.
The resulting debloated containers were then verified using all workloads to assess their generality.

The results show that 18 out of 20 containers were successfully verified, i.e., successfully executed all the workloads, indicating that debloating based on a subset of workloads can still allow the debloated containers to function with the full workloads set in most cases.
This is because debloating operates at the file level, where files essential for one workload may include functionality required for other workloads.
As a result, debloating at the file level tends to overestimate the resources necessary for the container to function correctly, ensuring a higher likelihood of generality across workloads.

For the remaining two failed containers, some workloads failed to work.
The first container, \texttt{ghost:7.4.1}, which has five workloads in total, was debloated using the default page-serving workload.
During verification, four out of the five workloads were successfully executed. The workload \texttt{sitemap}, which relies on the file \texttt{sitemap.xml}, failed because this file was removed during the debloating process.
The second container, \texttt{httpd:2.4}, was also debloated using the default page-serving workload.
It has four workloads in total, of which three were successfully verified.
The workloads \texttt{list-directory}, which lists the contents of a directory, failed because the directory was removed during debloating.
Notably, despite not passing all verification workloads, these two containers—debloated with a single basic workload—were still able to execute several other workloads.
This observation suggests that debloated containers, even when debloated using a minimal subset of workloads, can generalize to additional workloads to a certain extent.

To address the generality issue, we propose two potential solutions:
Firstly, we suggest that with a comprehensive set of debloating workloads, debloated containers could maintain full functionality across all required workloads, thereby improving their generality. One way to obtain such a comprehensive workload set is through online debloating, where containers are debloated in a production environment using real workloads rather than predefined workloads.
Secondly, we recommend selecting appropriate containers for debloating.
Container debloating is more suitable for containers with well-defined functionalities, such as serverless function containers.

The first approach requires that the debloater imposes minimal performance overhead on the target container.
To assess this, we evaluate the performance overhead of BAFFS in \S\ref{sec:performance-overhead}
Then we debloat serverless function containers to demonstrate the impact of debloating on serverless functions in \S\ref{sec:serverless}.

\subsection{Performance Overhead of BAFFS}\label{sec:performance-overhead}
We perform a filesystem benchmark on BAFFS to evaluate the performance overhead of BAFFS.
Specifically,  We conduct the disk benchmarks~\cite{DiskTest} using the Phoronix test suite~\cite{Phoronix}.
We first run the disk benchmarks in a container with the original filesystem.
Then we convert the filesystem into fully-sharing BAFFS and run the same disk benchmarks in the converted container.
The Flexible IO Tester of the disk benchmark was executed in the container.
Only read operations were measured as BAFFS does not affect the writing layer of a container.
Both random read and sequential read operations were measured with block sizes of 4KB and 2MB, and the bandwidth and I/O per second (IOPS) were recorded.
Then we compare the performance metrics of the original container with the container of BAFFS.
The performance metrics of the container of BAFFS were divided by the same metrics of the original container to obtain the relative overhead.

\begin{figure}
    \centering
    \includegraphics[scale=0.27]{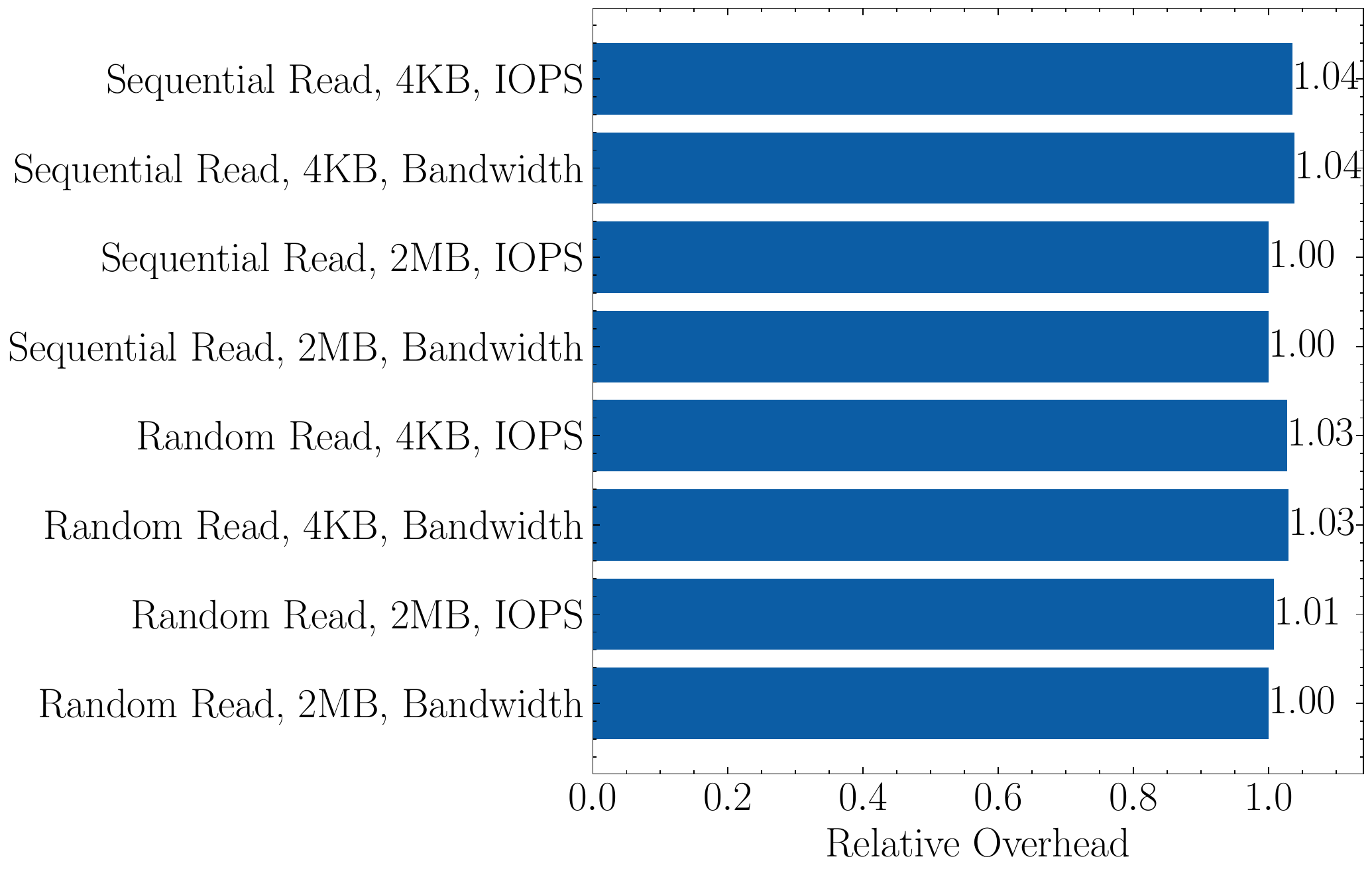}
    \caption{Relative overhead of file-system read operations. Lower is better.}
    \label{fig:phoronix_relative_overhead}
    \vspace{-2em}
\end{figure}

Figure \ref{fig:phoronix_relative_overhead} shows the relative overhead  (\textit{x-axis}) of the read operations of BAFFS.
The results show that all the metrics are around 1, indicating that the performance of BAFFS and the original container filesystem are similar.
The debloating layer does not incur much performance overhead.
Although the tested mode here is the fully-sharing BAFFS, it is also reasonable to assume that the no-sharing BAFFS could achieve the same performance since it has fewer debloating layers than the fully-sharing BAFFS.
The low performance overhead of BAFFS enables its use in online debloating, providing a practical solution to address the generality issue of container debloating.

\subsection{Debloat Serverless Function Containers}\label{sec:serverless}
We debloated the function containers of Serverless Benchmark~\cite{copik2021sebs}, which comprises 10 serverless functions.
The functions are packaged as containers.
One of the functions requiring special configuration on the serverless platform was excluded from the evaluation.
The remaining nine functions were debloated and evaluated.
We evaluate the cold start latency improvements by deploying the function containers on two serverless platforms: an open-source serverless platform, OpenWhisk~\cite{djemame2020open}, and a commercial serverless platform, AWS Lambda~\cite{amazonServerlessFunction}.

\subsubsection{Evaluation on OpenWhisk}
We first evaluated the performance improvement of the debloated serverless functions using OpenWhisk.
The framework was deployed on a machine with 16 CPUs and 64GB of memory in the standalone mode.
We use DockerHub as the container registry.
Two cold start scenarios were evaluated:
I. Cold start with the base image: The base image is already available on the serverless platform and only the layers unique to the target container need to be pulled from the registry.
II. Cold start without the base image: Both the function container and the base image must be pulled from the registry. This can occur in edge computing environments where the base image is not available on the edge device.

For scenario I, we applied the semi-sharing mode to debloat the nine containers.
For each original (debloated) container, we perform the cold start 50 times and record the cold start latency.
The results, summarized in Table~\ref{tab:serverless-improve}, include the container sizes and medians of the cold start latency of the original and debloated containers.
The container sizes were reduced by 1\% to 14\%, which is less significant compared to the reductions observed in previous experiments.
This is because the semi-sharing mode only debloats the unique layers of each container, while the base image layers remain unchanged for reusability.
Despite the modest size reductions, all functions exhibited improvements in cold start latency, ranging from 4\% to 25\%.
The standard deviation of cold start latency for all containers was less than 6\%.
The data also reveals that the larger the reduction in container size, the greater the reduction in cold start latency.

\begin{table}[htbp]
    \centering
    \caption{Scenario I and semi-sharing mode: Container size and cold start latency of original and debloated function containers. Percentages in parentheses are reductions. Ori.=Original, Deb.=Debloated.}
    \label{tab:serverless-improve}
    \begin{small}
        \begin{tabular}{lrrrr}
            \toprule
            \multirow{2}{*}{Function} & \multicolumn{2}{c}{Size/MB} & \multicolumn{2}{c}{Latency/ms}                                                       \\ \cmidrule(l{3pt}r{3pt}){2-3} \cmidrule(l{3pt}r{3pt}){4-5}
                                      & \multicolumn{1}{c}{Ori.}    & \multicolumn{1}{c}{Deb.}       & \multicolumn{1}{c}{Ori.} & \multicolumn{1}{c}{Deb.} \\
            \midrule
            dynamic-html              & 1,085 (1\%)                 & 1,079                          & 3,094 (4\%)              & 2,960                    \\
            uploader                  & 1,084 (1\%)                 & 1,078                          & 3,247 (4\%)              & 3,123                    \\
            thumbnailer               & 1,099 (1\%)                 & 1,084                          & 3,328 (7\%)              & 3,092                    \\
            video-proc.               & 1,339 (14\%)                & 1,154                          & 5,900 (25\%)             & 4,421                    \\
            compression               & 1,084 (1\%)                 & 1,078                          & 3,527 (4\%)              & 3,395                    \\
            image-recog.              & 1,802 (9\%)                 & 1,642                          & 14,519 (24\%)            & 10,999                   \\
            graph-page.               & 1,093 (1\%)                 & 1,087                          & 3,198 (5\%)              & 3,052                    \\
            graph-mst                 & 1,093 (1\%)                 & 1,087                          & 3,179 (5\%)              & 3,034                    \\
            graph-bfs                 & 1,093 (1\%)                 & 1,087                          & 3,195 (5\%)              & 3,035                    \\
            \bottomrule
        \end{tabular}
    \end{small}
\end{table}

For scenario II,
we performed the same experiments except that no-sharing mode is used to debloat the containers because the base image is not available.
As shown in Table~\ref{tab:serverless-improve-aggressive}, debloated by the no-sharing mode, the container sizes were significantly reduced, with reductions ranging from 66\% to 96\%.
Correspondingly, the cold start latencies also decreased substantially, with reductions between 46\% and 68\%.
The standard deviation of the cold start latency was less than 6\%.

\begin{table}[htbp]
    \centering
    \caption{Scenario II and no-sharing mode: Container size and cold start latency of original and debloated function containers. Percentages in parentheses are reductions. Ori.=Original, Deb.=Debloated.}
    \label{tab:serverless-improve-aggressive}
    \begin{small}
        \begin{tabular}{lrrrr}
            \toprule
            \multirow{2}{*}{Function} & \multicolumn{2}{c}{Size/MB} & \multicolumn{2}{c}{Latency/ms}                                                       \\ \cmidrule(l{3pt}r{3pt}){2-3} \cmidrule(l{3pt}r{3pt}){4-5}
                                      & \multicolumn{1}{c}{Ori.}    & \multicolumn{1}{c}{Deb.}       & \multicolumn{1}{c}{Ori.} & \multicolumn{1}{c}{Deb.} \\
            \midrule
            dynamic-html              & 1,085 (96\%)                & 49                             & 18,192 (68\%)            & 5,839                    \\
            uploader                  & 1,084 (95\%)                & 49                             & 18,682 (68\%)            & 6,029                    \\
            thumbnailer               & 1,099 (95\%)                & 55                             & 18,854 (68\%)            & 6,017                    \\
            video-proc.               & 1,339 (91\%)                & 125                            & 21,737 (67\%)            & 7,244                    \\
            compression               & 1,084 (96\%)                & 49                             & 19,206 (67\%)            & 6,335                    \\
            image-recog.              & 1,802 (66\%)                & 619                            & 28,283 (46\%)            & 15,392                   \\
            graph-page.               & 1,093 (95\%)                & 60                             & 18,619 (62\%)            & 7,158                    \\
            graph-mst                 & 1,093 (95\%)                & 60                             & 18,869 (67\%)            & 6,162                    \\
            graph-bfs                 & 1,093 (95\%)                & 60                             & 18,682 (68\%)            & 5,975                    \\
            \bottomrule
        \end{tabular}
    \end{small}
    \vspace{-1em}
\end{table}

Comparing Table~\ref{tab:serverless-improve} and Table~\ref{tab:serverless-improve-aggressive}, we observe that the absence of a base image in scenario II leads to a higher cold start latency for the original containers, compared to scenario I.
Furthermore, the \emph{debloated} cold start latency in scenario II is even higher than the \emph{original} cold start latency in scenario I.
This indicates that in scenario I, using the no-sharing mode for debloating can actually increase the cold start latency, contrary to expectations.
This is because when pulling the original container, the serverless framework benefits from the presence of the base image; it only needs to retrieve the unique layers of the target container, which are relatively small and can be pulled quickly.
In contrast, when pulling a debloated container created using the no-sharing mode, the serverless framework must download all the layers of the debloated container, resulting in a larger data transfer and slower performance.
We note that both Cimplifier and SlimToolKit have this issue.
However, the flexibility of BAFFS makes it able to address this issue by using the semi-sharing mode.

\begin{table}[H]
    \centering
    \caption{Container size and memory usage of function containers on AWS Lambda. The percentage in parentheses is the reduction.}
    \label{tab:serverless-improve-aws}
    \begin{small}
        \begin{tabular}{lrr}
            \toprule
            Function     & Container Size/MB & Memory Usage/MB \\
            \midrule
            dynamic-html & 537  (72\%)       & 40   (2\%)      \\
            uploader     & 535  (67\%)       & 79   (3\%)      \\
            thumbnailer  & 564  (68\%)       & 93   (2\%)      \\
            video-proc.  & 789  (68\%)       & 321  (0\%)      \\
            compression  & 535  (67\%)       & 101  (1\%)      \\
            image-recog. & 1,893( 61\%)      & 681  (1\%)      \\
            graph-page.  & 553  (72\%)       & 41   (2\%)      \\
            graph-mst    & 553  (72\%)       & 41   (2\%)      \\
            graph-bfs    & 553  (72\%)       & 41   (2\%)      \\
            \bottomrule
        \end{tabular}
    \end{small}
\end{table}

\subsubsection{Evaluation on AWS Lambda}
\begin{figure}[htbp]
    \centering
    \begin{subfigure}[b]{0.15\textwidth}
        \includegraphics[width=\textwidth]{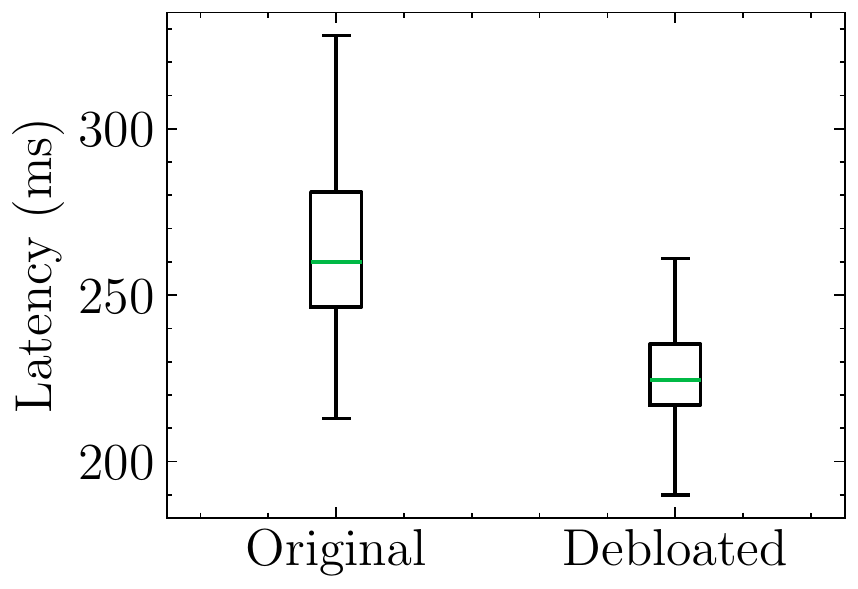}
        \caption{dynamic-html}
        \label{fig:billed-dynamic-html}
    \end{subfigure}
    \hfil
    \begin{subfigure}[b]{0.15\textwidth}
        \includegraphics[width=\textwidth]{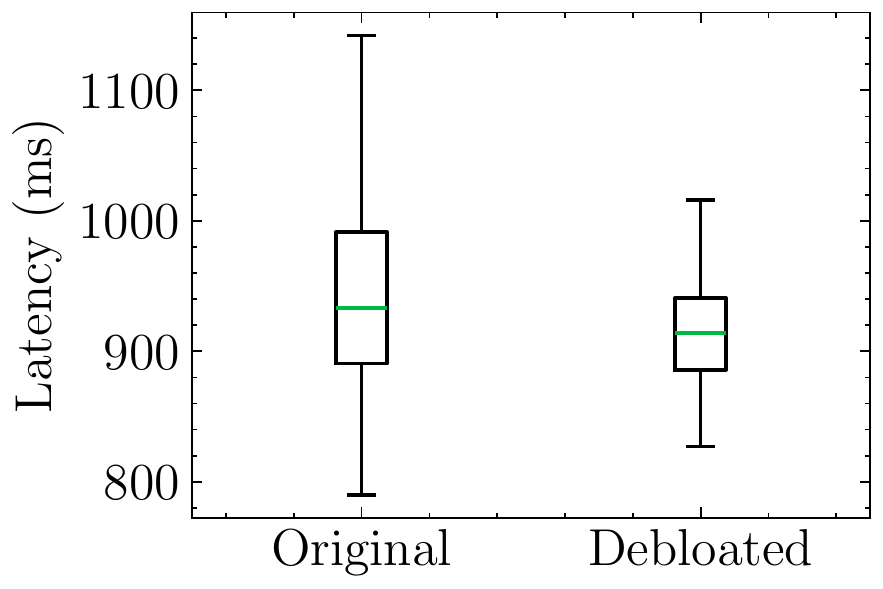}
        \caption{uploader}
        \label{fig:billed-uploader}
    \end{subfigure}
    \hfil
    \begin{subfigure}[b]{0.15\textwidth}
        \includegraphics[width=\textwidth]{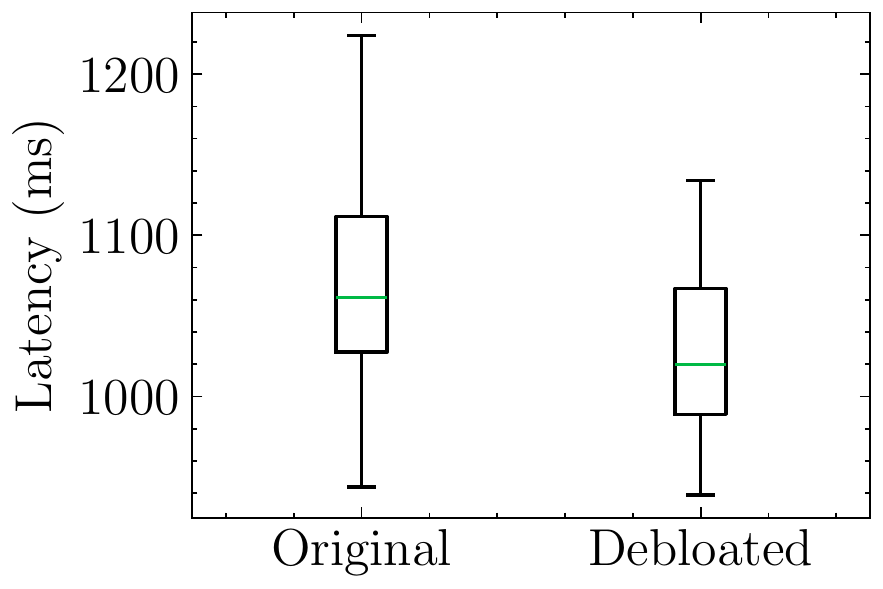}
        \caption{thumbnailer}
        \label{fig:billed-pulling_time}
    \end{subfigure}
    \vskip\baselineskip
    \begin{subfigure}[b]{0.15\textwidth}
        \includegraphics[width=\textwidth]{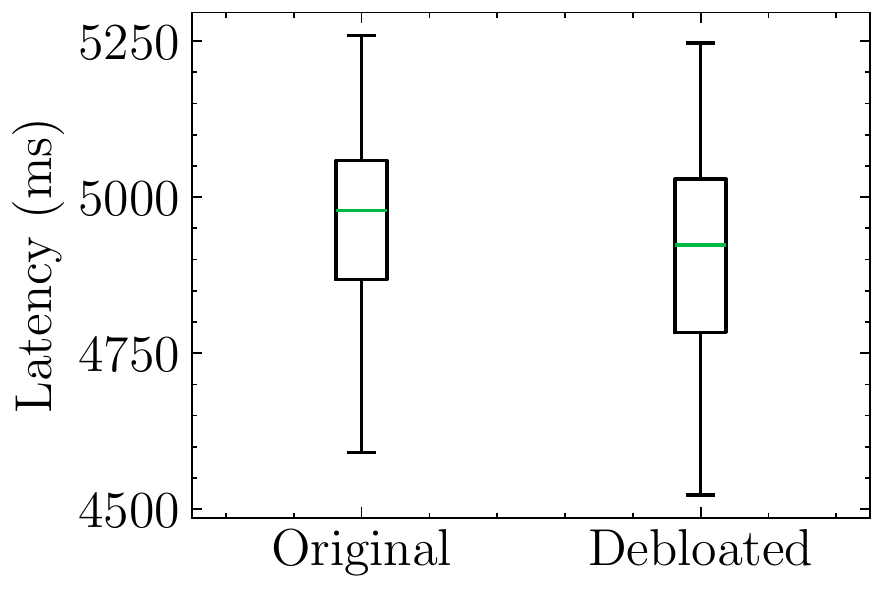}
        \caption{video-processing}
        \label{fig:billed-video-processing}
    \end{subfigure}
    \hfil
    \begin{subfigure}[b]{0.15\textwidth}
        \includegraphics[width=\textwidth]{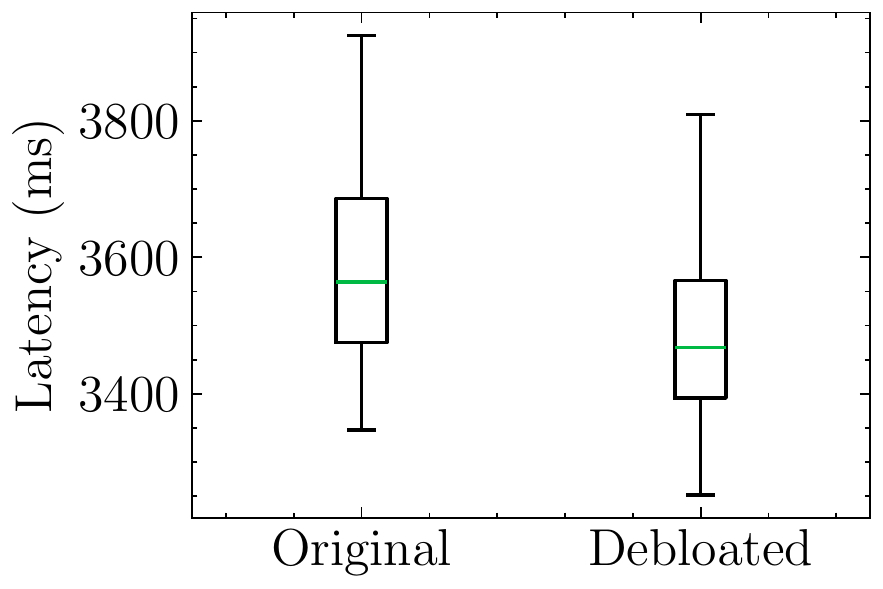}
        \caption{compression}
        \label{fig:billed-compression}
    \end{subfigure}
    \hfil
    \begin{subfigure}[b]{0.15\textwidth}
        \includegraphics[width=\textwidth]{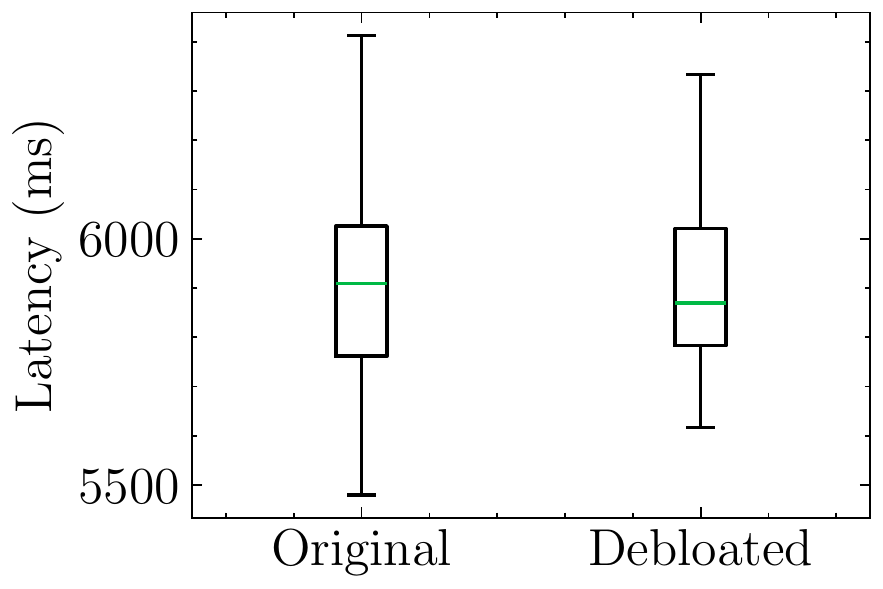}
        \caption{image-recognition}
        \label{fig:startup_time}
    \end{subfigure}
    \vskip\baselineskip
    \begin{subfigure}[b]{0.15\textwidth}
        \includegraphics[width=\textwidth]{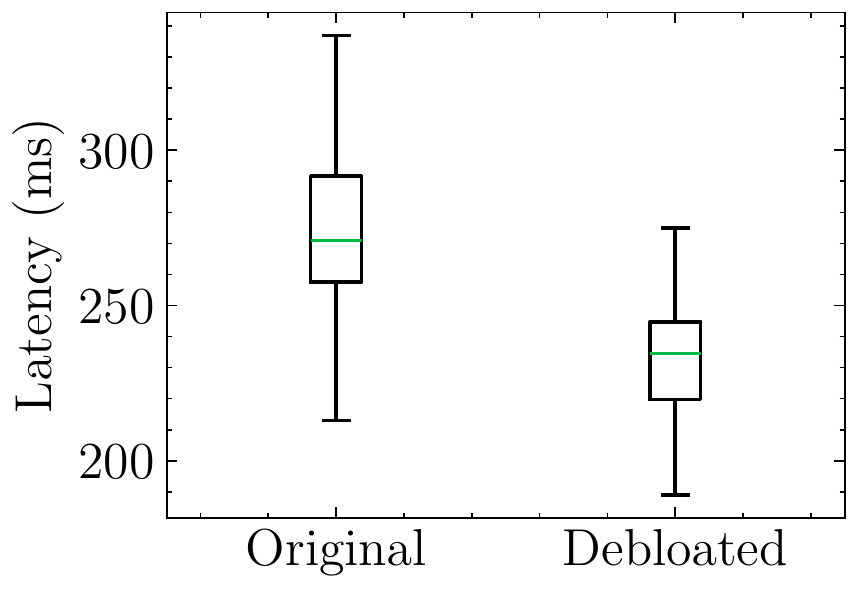}
        \caption{graph-pagerank}
        \label{fig:billed-pagerank}
    \end{subfigure}
    \hfil
    \begin{subfigure}[b]{0.15\textwidth}
        \includegraphics[width=\textwidth]{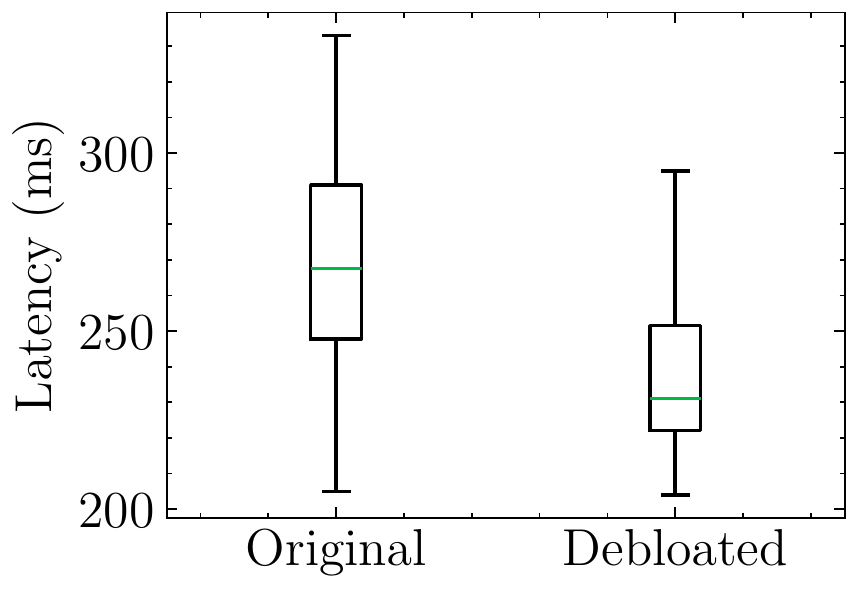}
        \caption{graph-mst}
        \label{fig:billed-graph-mst}
    \end{subfigure}
    \hfil
    \begin{subfigure}[b]{0.15\textwidth}
        \includegraphics[width=\textwidth]{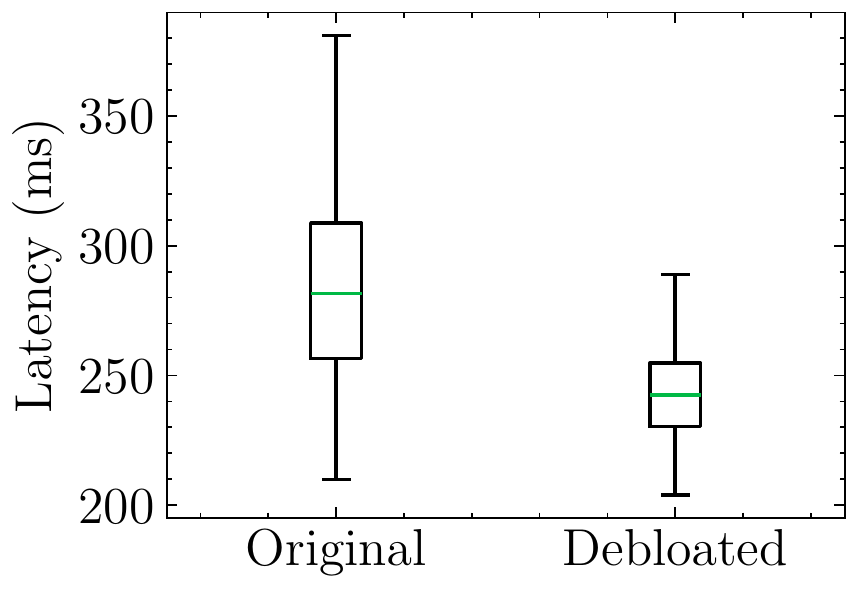}
        \caption{graph-bfs}
        \label{fig:billed-graph-bfs}
    \end{subfigure}
    \caption{Cold start latency of the original and debloated containers for each function on AWS Lambda.}
    \label{fig:aws-lambda-results}
    \vspace{-1em}
\end{figure}

We also performed the same evaluation on a commercial serverless platform, AWS Lambda.
Due to the proprietary nature of AWS Lambda's implementation details, we attempted debloating using both the semi-sharing and no-sharing modes to see which mode is more suitable for AWS Lambda.
Our findings show that the no-sharing mode is more suitable for AWS Lambda, achieving better performance improvements, while the semi-sharing mode did not show significant improvements.
Therefore, we present only the results of the no-sharing mode in Table~\ref{tab:serverless-improve-aws}.
The evaluation focused on two metrics provided by the AWS Lambda platform: memory usage and cold start latency.
Memory usage measures the amount of memory consumed by the function during execution.
We use the billed duration reported by the platform as the cold start latency metric, which represents the time from when the function begins executing until it terminates, rounded up to the nearest millisecond.
x
Table~\ref{tab:serverless-improve-aws} shows the original container size, memory usage and their reductions after debloating.
For AWS Lambda  function containers, the container sizes were reduced by 61\% to 72\%.
Besides the container size reduction, the memory usage was also reduced by 0\% to 3\%.
Figure~\ref{fig:aws-lambda-results} shows the cold start latency of the original and debloated containers for each function.
For all functions, the debloated containers showed lower cold start latency compared to the original containers.
We calculate the relative improvement of the cold start latency using the median values.
The functions \texttt{dynamic-html}, \texttt{graph-mst} and \texttt{graph-bfs} achieved the highest improvements, with reductions of 14\% in cold start latency.
The minimal improvement of 1\% was observed for the \texttt{video-processing} and \texttt{image-recognition}.
AWS Lambda charges are based on execution latency, so reducing cold start latency directly translates to cost savings.
Additionally, since AWS Lambda hosts containers in its registry, smaller container sizes can further reduce costs associated with storage and network bandwidth.


\subsection{Evaluation on Lazy-Loading Snapshotters}
Lazy loading snapshotters are used to improve container provisioning time.
BAFFS can be combined with lazy-loading snapshotters to further enhance provision performance.
Lazy-loading snapshotters first convert container images into a lazy-loading format, a process that can be time-consuming~\cite{chen2022starlight}.
This conversion latency can lead to delays in updating serverless functions.
Once converted, the image is pushed to and stored in a registry.
During container startup, only the  essential files for startup are immediately pulled, while the remaining files are fetched in the background.
This mechanism allows the container to start faster.
In this section, we evaluate whether debloating can improve two key metrics of lazy-loading snapshotters:(1) Image format conversion time and (2) Container pull time.
Two lazy-loading snapshotters, eStargz and Starlight, are evaluated.
First, we reused the same nine OpenWhisk function containers and debloated them using the no-sharing mode.
The semi-sharing or full-sharing modes were not used, as lazy-loading snapshotters do not rely on the base image to start the container.
Then we converted both the original and debloated containers into a lazy-loading format using eStargz and Starlight.

The time required to convert each container was measured. Table~\ref{tab:snapshotter-convert-time} presents the results.
Both eStargz and Starlight exhibited significant reductions in conversion time for debloated containers.
For eStargz, the conversion time was reduced by 28\% to 79\%.
For Starlight, the reduction ranged from 45\% to 93\%.
The standard deviation of the conversion time was less than 3\%.
These results demonstrate that debloating significantly reduces the time needed to convert container images into a lazy-loading format.

\begin{table}[htbp]
    \centering
    \caption{The conversion time of eStargz and Starlight. Percentages in parentheses are reductions. Time is in seconds.}
    \label{tab:snapshotter-convert-time}
    \begin{small}
        \begin{tabular}{lrr}
            \toprule
            Function     & eStargz     & Starlight   \\
            \midrule
            dynamic-html & 73s  (79\%) & 114s (93\%) \\
            uploader     & 67s  (77\%) & 113s (93\%) \\
            thumbnailer  & 68s  (77\%) & 114s (93\%) \\
            video-proc.  & 70s  (70\%) & 115s (92\%) \\
            compression  & 67s  (77\%) & 114s (93\%) \\
            image-recog. & 111s (28\%) & 117s (45\%) \\
            graph-page.  & 68s (77\%)  & 113s (93\%) \\
            graph-mst    & 63s  (76\%) & 112s (93\%) \\
            graph-bfs    & 65s  (76\%) & 114s (93\%) \\
            \bottomrule
        \end{tabular}
    \end{small}
    \vspace{-1em}
\end{table}

\begin{figure}[htbp]
    \centering
    \begin{subfigure}[b]{0.15\textwidth}
        \includegraphics[width=\textwidth]{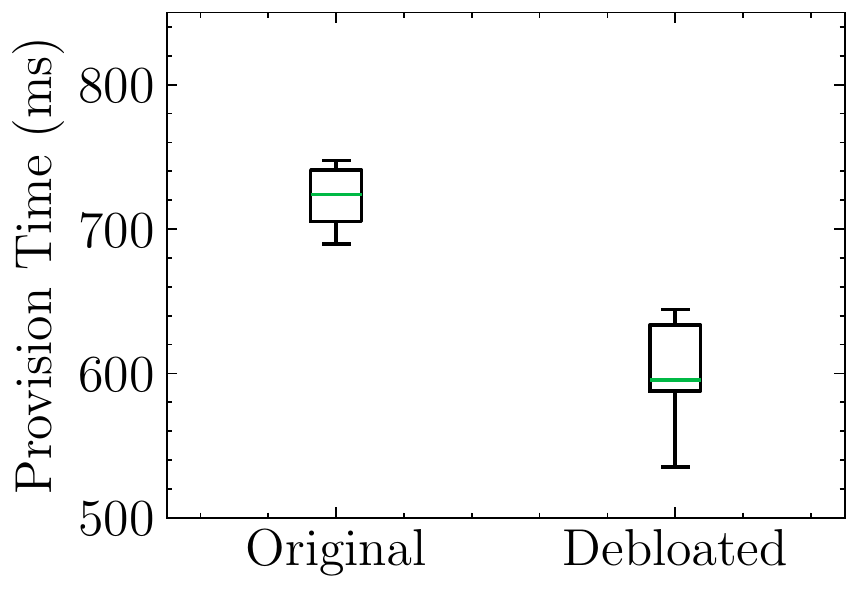}
        \caption{dynamic-html}
        \label{fig:provision-dynamic-html}
    \end{subfigure}
    \hfil
    \begin{subfigure}[b]{0.15\textwidth}
        \includegraphics[width=\textwidth]{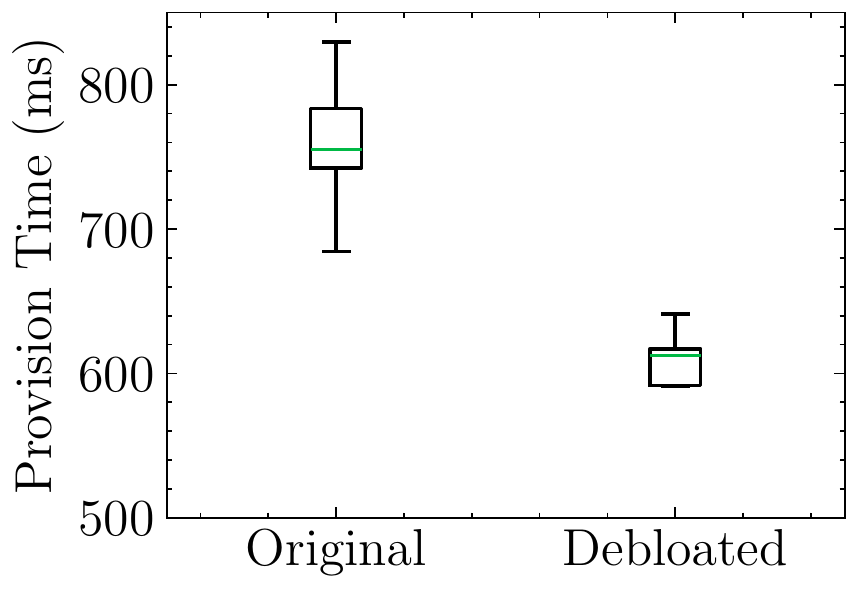}
        \caption{uploader}
        \label{fig:provision-uploader}
    \end{subfigure}
    \hfil
    \begin{subfigure}[b]{0.15\textwidth}
        \includegraphics[width=\textwidth]{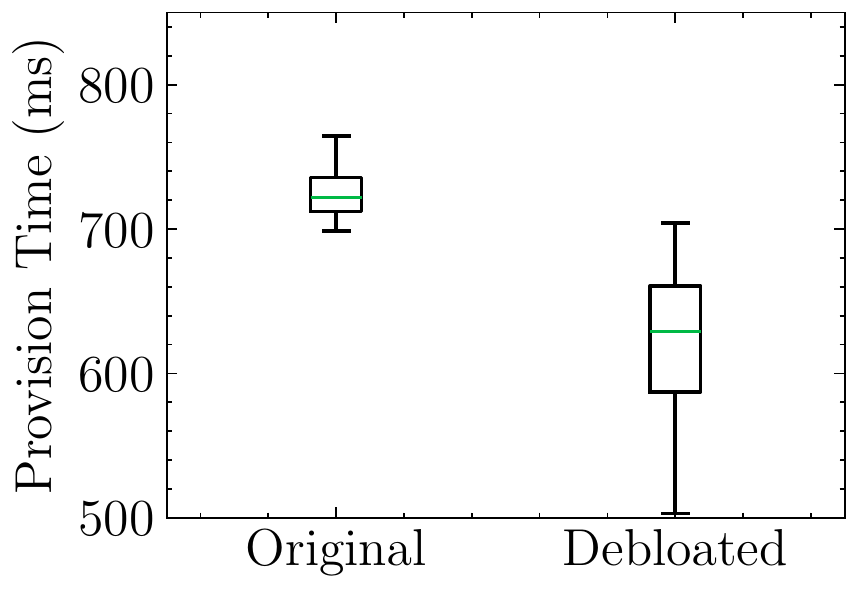}
        \caption{thumbnailer}
        \label{fig:provision-tjumbnailer}
    \end{subfigure}
    \vskip\baselineskip
    \begin{subfigure}[b]{0.15\textwidth}
        \includegraphics[width=\textwidth]{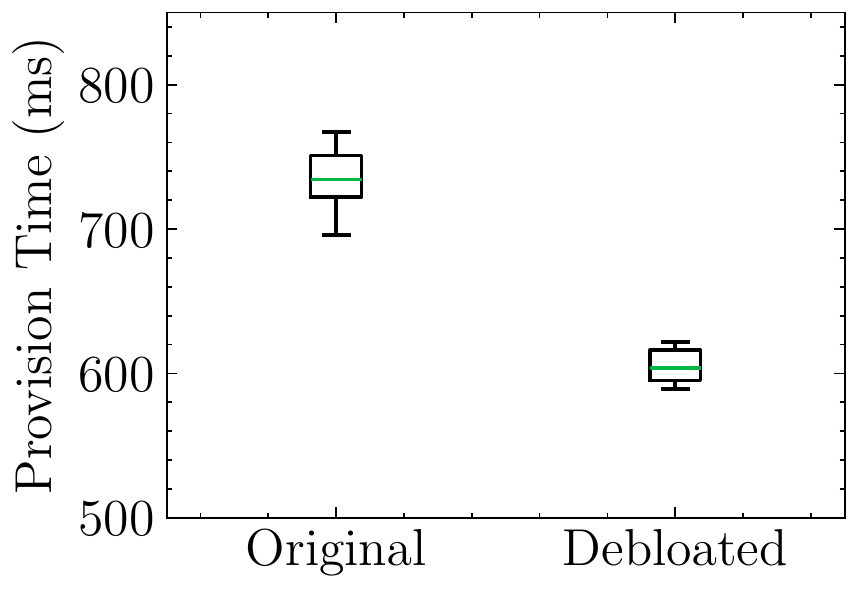}
        \caption{video-processing}
        \label{fig:provision-video-processing}
    \end{subfigure}
    \hfil
    \begin{subfigure}[b]{0.15\textwidth}
        \includegraphics[width=\textwidth]{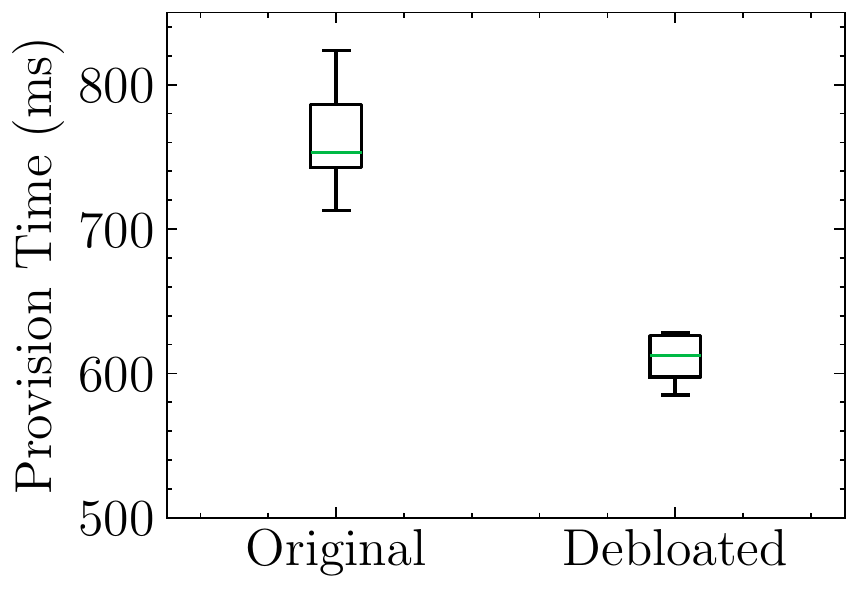}
        \caption{compression}
        \label{fig:provision-compression}
    \end{subfigure}
    \hfil
    \begin{subfigure}[b]{0.15\textwidth}
        \includegraphics[width=\textwidth]{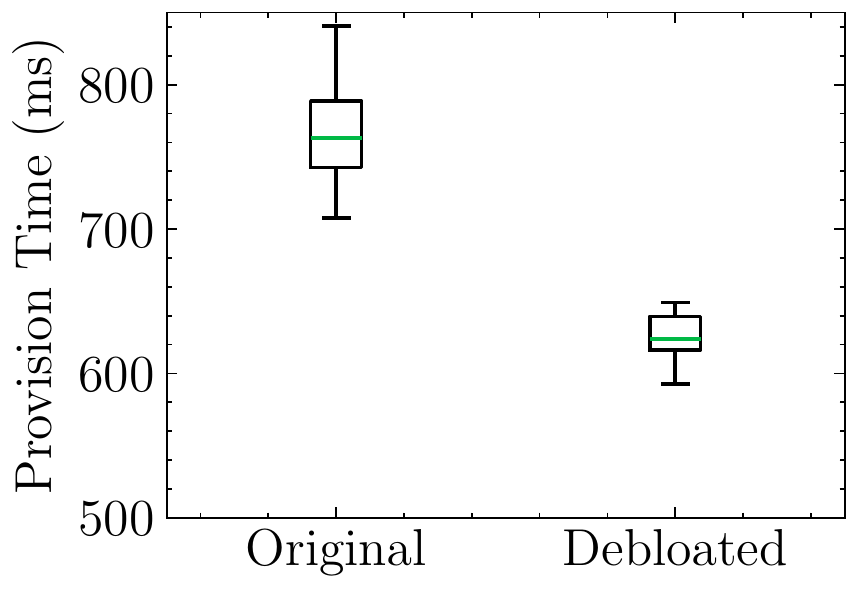}
        \caption{Startup time.}
        \label{fig:provision-image-recognition}
    \end{subfigure}
    \vskip\baselineskip
    \begin{subfigure}[b]{0.15\textwidth}
        \includegraphics[width=\textwidth]{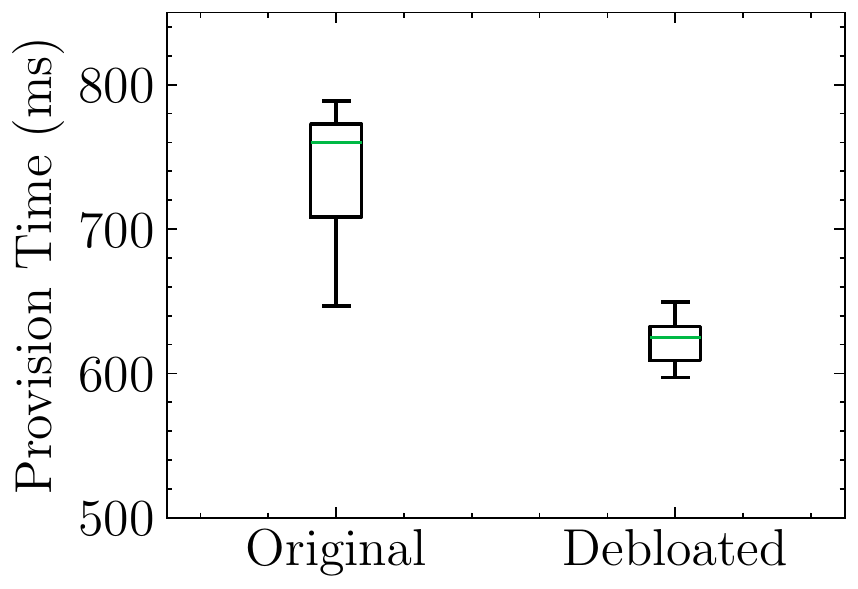}
        \caption{graph-pagerank}
        \label{fig:provision-pagerank}
    \end{subfigure}
    \hfil
    \begin{subfigure}[b]{0.15\textwidth}
        \includegraphics[width=\textwidth]{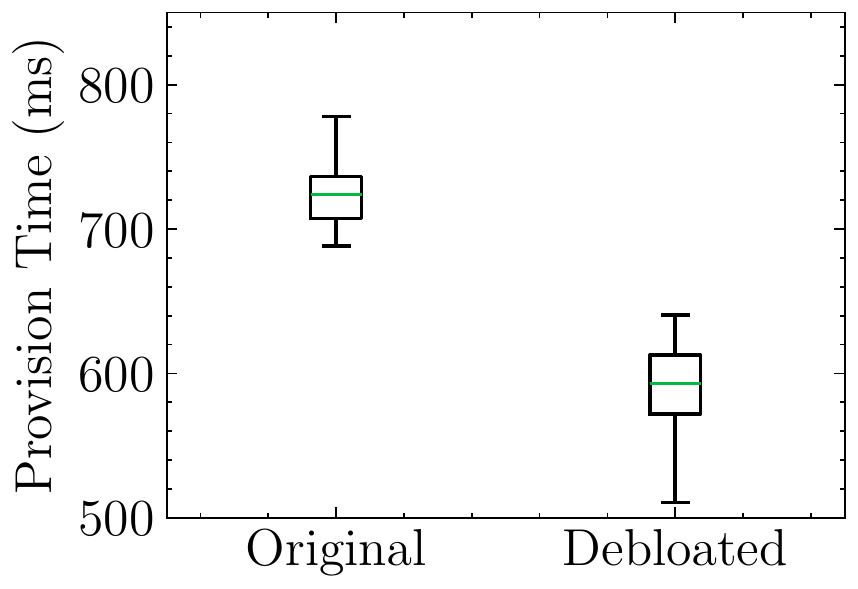}
        \caption{graph-mst}
        \label{fig:provision-graph-mst}
    \end{subfigure}
    \hfil
    \begin{subfigure}[b]{0.15\textwidth}
        \includegraphics[width=\textwidth]{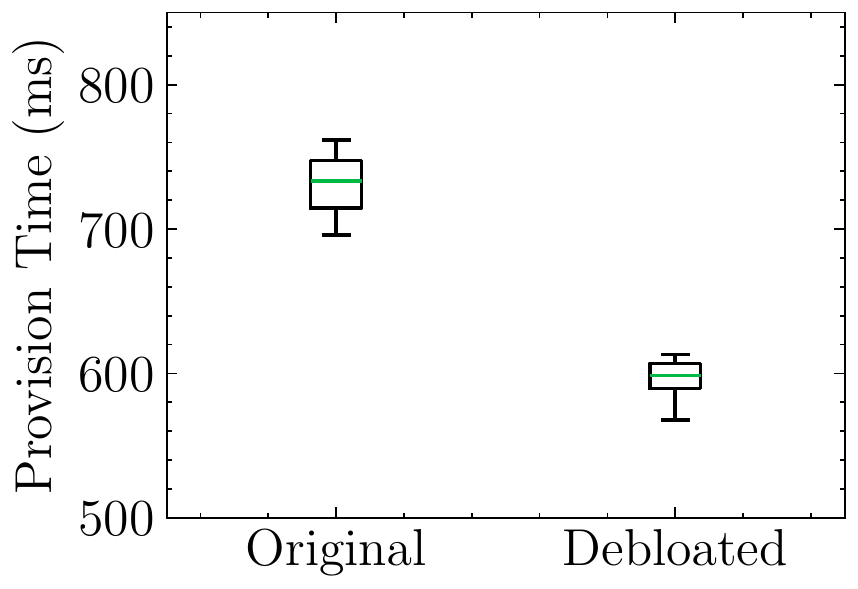}
        \caption{graph-bfs}
        \label{fig:provision-graph-bfs}
    \end{subfigure}
    \caption{Provision time of the original and debloated containers for each function using eStargz.}
    \label{fig:esgz-provision-results}
    \vspace{-1em}
\end{figure}

Figure~\ref{fig:esgz-provision-results} presents the provisioning times for both the original and debloated containers.
Starlight was excluded from this evaluation, as it failed to provision either the original or debloated containers.\footnote{An issue has been submitted to the Starlight repository.}
The results show that with eStargz, the provisioning times for debloated containers were reduced by 13\% to 19\%.
These findings confirm that container debloating effectively complements lazy-loading snapshotters, reducing the time required for format conversion and container provisioning, and thereby improving the overall performance of lazy-loading snapshotters.
We note that the AWS Lambda function containers had similar results.

\subsection{Evaluation of Mode Selection Strategy}\label{sec:layer-sharing}
\begin{figure}
    \centering
    \includegraphics[scale=0.23]{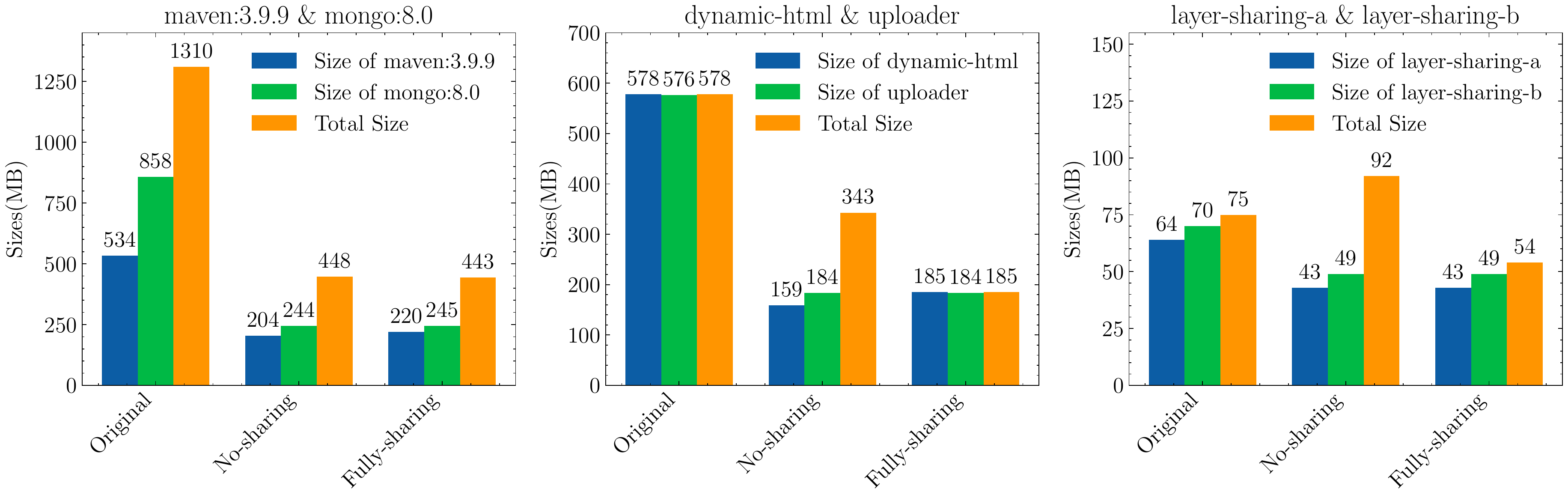}
    \caption{Container sizes under different modes.}
    \label{fig:layer_sharing_results}
    \vspace{-1.36em}
\end{figure}

To evaluate the mode selection strategy of BAFFS,
we selected containers with shared layers, including two containers \texttt{maven:3.9.9} and \texttt{mongo:8.0} from DockerHub, alongside AWS Lambda function containers \texttt{dynamic-html} and \texttt{uploader}.
Additionally, we incorporated \texttt{layer-sharing-a} and \texttt{layer-sharing-b}, two manually created containers with shared layers, to further our investigation.
We debloated these containers using no-sharing and fully-sharing BAFFS. The results are presented in Figure \ref{fig:layer_sharing_results}.

For containers \texttt{maven:3.9.9} and \texttt{mongo:8.0}, originally sized at 534MB and 858MB (totaling 1,310MB due to shared layers),
debloating with no-sharing BAFFS resulted in a significant size reduction, bringing the total size down to 448MB.
When debloated using fully-sharing BAFFS, while the total size is smaller, individual container sizes are larger than their no-sharing counterparts.
The $\theta$ value of these two containers is 0.3.
This means that debloating these two containers using fully-sharing BAFFS will cause each container to include more unneeded files.
Therefore, no-sharing BAFFS for these containers is more suitable.

In the case of containers \texttt{dynamic-html} and \texttt{uploader}, with original sizes of 578MB and 576MB (totaling 578MB),
no-sharing BAFFS reduced the total size to 343MB.
Notably, layer-sharing BAFFS markedly decreases the total size to 185MB.
This significant size reduction, supported by a $\theta$ value of 6, indicates a clear advantage of fully-sharing BAFFS for these containers.

The experiment with \texttt{layer-sharing-a} and \texttt{layer-sharing-b} reveals an intriguing aspect of container debloating.
The original total size of two containers is 75MB, debloating with no-sharing BAFFS paradoxically increases their total size to 92MB.
This issue, that the total size of debloated containers can exceed that of the original ones, is both faced by Cimplifier and SlimToolKit.
The reason for this issue is that these tools break the layer-sharing feature of container filesystems.
However, fully-sharing BAFFS utilizes the layer-sharing feature and effectively reduces total sizes to 54MB.
A remarkably high $\theta$ value of 380,000 in this scenario indicates that fully-sharing BAFFS is more suitable.

\section{Discussion}

Software bloat in containers leads to resource inefficiencies and prolonged provisioning times.
In this paper, we introduced BAFFS, a flexible filesystem designed to reduce software bloat in container images.
BAFFS leverages the layered filesystem architecture of containers to reduce bloat.
BAFFS has a basic building block: the debloating layer, which can be used to create different architectures and provide greater flexibility.
Our results demonstrate that BAFFS effectively reduces container bloat, improves serverless function cold-start times, and shortens container provisioning times.

As containers have become the standard for software packaging in cloud computing, they offer great convenience but at the expense of efficiency.
Although many techniques have been proposed to improve the efficiency of containers, they only solve the symptoms of the problem, but not the root cause.
We believe that the root cause of these inefficiencies, such as long provisioning time and increased resource usage, is container bloat.
Container debloating, as demonstrated by BAFFS, provides a method to balance this trade-off by improving container efficiency without sacrificing convenience.
Furthermore, BAFFS is orthogonal to these optimization techniques and can be combined with them to provide additional improvements.

Container debloating often faces challenges in terms of generality~\cite{xin2022studying}, as the workloads used for debloating may differ from those encountered in production, potentially causing crashes in the debloated container.
Two approaches can help mitigate this challenge:
Firstly, use more comprehensive workloads to debloat containers, such as performing online debloating with production workloads.
The low overhead of BAFFS makes it well-suited for online debloating.
Secondly, select the containers that have well-defined workloads.
The specific and well-defined nature of serverless functions makes it an ideal use case for debloating.




We tried to mitigate validity threats via our experimental design.
We referred to the official documentation of each container to identify its representative workloads, aiming to cover as many use cases as possible.
Furthermore, we have published our source code of the containers and corresponding workloads to the community to facilitate further improvement and research in container debloating.

\section{Related Work}

The problem of container bloat has been drawing attention from both academia and industry.
Apart from the lazy-loading snapshotters discussed in \S\ref{sec:background}, many other techniques have been proposed to address the effects of container bloat.
CNTR~\cite{thalheim2018cntr} introduces the concept of a slim and a fat container image. The three main use cases for CNTR are; Container to container debugging in production; Host to container debugging; and Container to host administration. Both the slim and fat containers run on the same host, and there are no space or bandwidth savings. 
Slacker~\cite{harter2016slacker} is a Docker storage driver designed to optimize fast container startup and reduce the time it takes to provision a container.
It provisions the container quickly using backend clones and minimizes startup latency by lazily fetching container data.
DADI~\cite{li2020dadi} is a block-level image service for increased agility and elasticity in deploying applications by providing fine-grained on-demand transfer of remote images.
FAASNET~\cite{wang2021faasnet} is a middle-ware system designed for highly scalable container provisioning in serverless platforms, which enables scalable container provisioning via a lightweight, adaptive function tree structure and uses an on-demand fetching mechanism to reduce provisioning costs.
Gear~\cite{fan2021gear} is a new image format that reduces container deployment time and storage size of the image registry by separating the index that describes the filesystem structure from the files that are required for running an application.
We argue that these optimizations only solve the symptoms of the problem, but not the root cause.
We believe that the root cause of long provisioning time and increased resource usage is container bloat, and BAFFS can be combined with them to provide further improvements.
\section{Conclusion}

This paper addresses the issue of software bloat in containers, which impacts provisioning times, resource utilization, and overall system performance.
Through a large-scale analysis of the top 20 most downloaded containers on DockerHub, we demonstrate that container bloat is widespread, with over 50\% of containers containing more than 60\% of bloat.
Existing debloating tools have several inherent limitations, such as breaking the layered structure of container filesystems.
To overcome these limitations, we introduced BAFFS, a Bloat-Aware Flexible Filesystem that preserves container functionality while significantly reducing container bloat.
BAFFS supports multiple modes, including no-sharing, fully-sharing, and semi-sharing modes, making it adaptable to diverse use cases.
Our evaluations show that BAFFS reduces cold start latency of serverless functions by up to 68\% and, when integrated with lazy-loading snapshotters, enhances container provisioning performance by reducing conversion times by 93\% and provisioning times by 19\%.
BAFFS provides an effective and flexible solution to container debloating that balances the trade-off between container convenience and efficiency.

\newpage
\bibliographystyle{plain}
\bibliography{sample-base}
\appendix
\section{Appendix}
\subsection{Containers Evaluated}
\begin{table}[H]
    \centering
    \begin{threeparttable}
        \caption{Details of containers evaluated.}
        \label{tab:containers}
        \begin{small}
            \begin{tabular}{lrr}
                \toprule
                Container                & \#Workloads\tnote{1} & Pull Times \\
                \midrule
                httpd:2.4                & 4                    & 1B+        \\
                nginx:1.27.2             & 4                    & 1B+        \\
                memcached:1.6.32         & 6                    & 500M+      \\
                mysql:9.1                & 7                    & 1B+        \\
                postgres:17              & 4                    & 1B+        \\
                ghost:5.101.3            & 5                    & 500M+      \\
                redis:7.4.1              & 4                    & 1B+        \\
                haproxy:3.0.6            & 4                    & 1B+        \\
                mongo:8.0                & 7                    & 1B+        \\
                solr:9.7.0               & 4                    & 100M+      \\
                rabbitmq:4.0             & 4                    & 1B+        \\
                maven:3.9.9              & 3                    & 500M+      \\
                elasticsearch:8.16.0     & 4                    & 500M+      \\
                eclipse-mosquitto:2.0.20 & 5                    & 500M+      \\
                telegraf:1.30            & 3                    & 500M+      \\
                nextcloud:28.0.12        & 2                    & 500M+      \\
                sonarqube:9.9.7          & 3                    & 1B+        \\
                registry:2.8.3           & 4                    & 1B+        \\
                consul:1.15.4            & 4                    & 1B+        \\
                traefik:v3.2.0           & 3                    & 1B+        \\
                \bottomrule
            \end{tabular}
            \begin{tablenotes}
                \item[1] We categorized the features into workloads. Each workload may involve multiple features. For example, the \texttt{test\_index\_operations} workload for the \texttt{mysql:9.1} involves creating a table, inserting data, creating an index, and querying the data. For the details of the workloads, please refer \url{https://drive.google.com/drive/folders/1KJcy1sM2ytQF2YzLZKqU_sZsfO6Q7Glm}
            \end{tablenotes}
        \end{small}
    \end{threeparttable}
\end{table}

\subsection{Cimplifier Debloating Results}
\begin{table}[H]
    \centering
    \begin{threeparttable}
        \caption{Cimplifier Debloating Results of the top 20 most downloaded containers from DockerHub.
            Containers failed to debloat are not included in the table.
            Results are sorted by the reduction percentage.}
        \label{tab:cimplifier_all_size_reduction}
        \begin{small}
            \begin{tabular}{lrrr}
                \toprule
                Container                & \makecell{Original              \\(MB)} & \makecell{Debloated\\(MB)} & Reduction \\
                \midrule
                httpd:2.4        & 141                & 7   & 95\% \\
                nginx:1.27.2             & 183                & 12  & 93\% \\
                eclipse-mosquitto:2.0.20 & 14                 & 7   & 51\% \\
                telegraf:1.30            & 435                & 223 & 49\% \\
                nextcloud:28.0.12        & 1,200              & 761 & 37\% \\
                registry:2.8.3           & 24                 & 18  & 25\% \\
                traefik:v3.2.0           & 176                & 169 & 4\%  \\
                \bottomrule
            \end{tabular}
        \end{small}
    \end{threeparttable}
\end{table}

\subsection{SlimToolKit Debloating Results}
\begin{table}[H]
    \centering
    \begin{threeparttable}
        \caption{SlimToolKit Debloating Results of the top 20 most downloaded containers from DockerHub.
            Containers failed to debloat are not included in the table.
            Results are sorted by the reduction percentage.}
        \label{tab:slimtool_all_size_reduction}
        \begin{small}
            \begin{tabular}{lrrr}
                \toprule
                Container                & \makecell{Original              \\(MB)} & \makecell{Debloated\\(MB)} & Reduction \\
                \midrule
                nginx:1.27.2             & 183                & 13  & 93\% \\
                memcached:1.6.32         & 81                 & 9   & 89\% \\
                haproxy:3.0.6            & 98                 & 27  & 72\% \\
                maven:3.9.9              & 505                & 199 & 61\% \\
                telegraf:1.30            & 435                & 223 & 49\% \\
                eclipse-mosquitto:2.0.20 & 14                 & 7   & 49\% \\
                registry:2.8.3           & 24                 & 19  & 24\% \\
                traefik:v3.2.0           & 176                & 170 & 4\%  \\
                \bottomrule
            \end{tabular}
        \end{small}
    \end{threeparttable}
\end{table}

\subsection{BAFFS Debloating Results}
\begin{table}[H]
    \centering
    \begin{threeparttable}
        \caption{BAFFS Debloating Results of the top 20 most downloaded containers from DockerHub. Results are sorted by the reduction percentage.}
        \label{tab:all_size_reduction}
        \begin{small}
            \begin{tabular}{lrrr}
                \toprule
                Container                & \makecell{Original              \\(MB)} & \makecell{Debloated\\(MB)} & Reduction \\
                \midrule
                httpd:2.4                & 141                & 7   & 95\% \\
                nginx:1.27.2             & 183                & 12  & 93\% \\
                memcached:1.6.32         & 81                 & 9   & 89\% \\
                mysql:9.1                & 574                & 99  & 83\% \\
                postgres:17              & 415                & 85  & 79\% \\
                ghost:5.101.3            & 547                & 121 & 78\% \\
                redis:7.4.1              & 112                & 27  & 75\% \\
                haproxy:3.0.6            & 98                 & 27  & 72\% \\
                mongo:8.0                & 815                & 233 & 71\% \\
                solr:9.7.0               & 561                & 195 & 65\% \\
                rabbitmq:4.0             & 209                & 73  & 65\% \\
                maven:3.9.9              & 505                & 195 & 61\% \\
                elasticsearch:8.16.0     & 1,241              & 479 & 61\% \\
                eclipse-mosquitto:2.0.20 & 14                 & 7   & 51\% \\
                telegraf:1.30            & 435                & 223 & 49\% \\
                nextcloud:28.0.12        & 1,200              & 761 & 37\% \\
                sonarqube:9.9.7          & 576                & 428 & 26\% \\
                registry:2.8.3           & 24                 & 18  & 25\% \\
                consul:1.15.4            & 148                & 137 & 7\%  \\
                traefik:v3.2.0           & 176                & 169 & 4\%  \\
                \bottomrule
            \end{tabular}
        \end{small}
    \end{threeparttable}
\end{table}

\end{document}